\documentclass[a4paper,fleqn,usenatbib]{mnras}

\usepackage{newtxtext,newtxmath}

\usepackage[T1]{fontenc}
\usepackage{ae,aecompl}

\usepackage{graphicx}	
\usepackage{amsmath}	
\usepackage{longtable}
\usepackage{hyperref}
\usepackage{breakurl}
\usepackage[normalem]{ulem}

\title[Multiwavelength monitoring of NGC 1275]{Multiwavelength monitoring of NGC 1275 
over a decade: Evidence of a shift in synchrotron peak frequency and long-term 
multi-band flux increase}

\author[S. Gulati et al.]{
Sanna Gulati,$^{1}$
Debbijoy Bhattacharya,$^{1}$ \thanks{E-mail: debbijoy.b@manipal.edu}
Subir Bhattacharyya$^{2,3}$
\newauthor Nilay Bhatt$^{2}$, C. S. Stalin$^{4}$ and V. K. Agrawal$^{5}$ 
\\
$^{1}$Manipal Centre for Natural Sciences, Centre of Excellence, Manipal Academy of Higher Education, Manipal - 576104, India\\
$^{2}$Bhabha Atomic Research Centre, Mumbai - 400085, India\\
$^{3}$Homi Bhabha National Institute, Anushaktinagar, Mumbai - 400094, India \\
$^{4}$Indian Institute of Astrophysics, Bangalore - 560034, India\\
$^{5}$Space Astronomy Group, U R Rao Satellite Centre, Bangalore - 560017, India \\}

\date{Accepted XXX. Received YYY; in original form ZZZ}

\pubyear{2020}

\begin{document}
\label{firstpage}
\pagerange{\pageref{firstpage}--\pageref{lastpage}}
\maketitle

\begin{abstract}
We carried out a
detailed study of the temporal and broadband 
spectral behaviour of one of 
the brightest misaligned 
active galaxies in $\gamma$-rays, 
NGC 1275 utilising
$11$ years of {\sl Fermi}, and available 
{\sl Swift} and {\sl AstroSat} observations. 
Based on the cumulative flux distribution of  
the $\gamma$-ray lightcurve, 
we identified four distinct activity states 
and noticed an increase in the baseline 
flux during the first three states. 
Similar nature of the increase in the average flux 
was also noticed in X-ray and UV bands. 
A large flaring activity in $\gamma$-rays 
was noticed in
the fourth state. The source was observed 
twice by {\sl AstroSat} for shorter intervals ($\sim$days) during the longer observing periods ($\sim$years) state 3 and 4. 
During {\sl AstroSat} observing periods, the source $\gamma$-ray flux was higher than the 
average flux observed during longer duration states. 
The increase in the average baseline flux 
from state 1 to state 3 can be explained
considering a corresponding increase of jet 
particle normalisation. 
The inverse Comptonisation of synchrotron photons 
explained the average X-ray and
$\gamma$-ray emission
by jet electrons during the first three longer duration states. 
However, during the shorter duration {\sl AstroSat} observing periods, 
a shift of the synchrotron peak frequency 
was noticed, and the synchrotron emission of jet electrons 
well explained the observed X-ray flux.

\end{abstract}




\section{Introduction}
NGC 1275 is one of the brightest nearby radio galaxies 
(z$=0.0176$; \citealt{falco1999}) 
situated at the centre of the Perseus cluster. The optical 
spectra of this galaxy exhibit strong emission lines which are a 
typical feature of Seyfert galaxies \citep{humason1932,khachikian1974}.  
However, in the radio band, this 
source is classified as a Fanaroff-Riley type I 
(FR I type) radio galaxy with 
a compact central source and 
an extended jet \citep[e.g.,][]{Vermeulen1994,Asada2006,buttiglione2010}.

NGC 1275 (4FGL J$0319.8+4130$) is one of the brightest misaligned 
active galaxies in $\gamma$-rays. 
\citet{ngc1275_fermi} reported the discovery of high energy $\gamma$-ray emission 
from NGC 1275 utilising the first few months of observations 
from the Large Area Telescope (LAT) onboard the {\sl Fermi} 
$\gamma$-ray space telescope ({\sl Fermi}).
 
Evidence of 
variability, in both, long \citep{kataoka2010, dutson2014}  
as well as short timescales with large flaring activities 
\citep{atel_2010,ngc1275_fermi_2011, atel_2013, atel_2015, Baghmanyan2017, kushwaha2017, 
 Tanada2018,Chitnis2020, Bitan2020} 
in $\gamma$-rays was noticed in this source.
NGC 1275 was also detected 
by Major Atmospheric Gamma Imaging Cherenkov ({\sl MAGIC}) 
telescope and Very Energetic Radiation 
Imaging Telescope Array System ({\sl VERITAS}) 
in very high energy $\gamma$-rays 
\citep{aleksic2012,Aleksic2014,benbow2015, mirzoyan2016, 
mirzoyan2017,mukherjeeveritas2016,mukherjeeveritas2017,magic2018}. 
Though earlier {\sl MAGIC} observations showed marginal flux variation 
in monthly timescales, \citet{magic2018} reported 
a presence of significant variation in ``day-by-day'' 
$\gamma$-ray lightcurve.

NGC 1275 was studied in hard X-ray band using observations from 
Nuclear Spectroscopic Telescope Array - {\sl NuSTAR} \citep{Tanada2018,Rani2018,Chitnis2020}.  
\citet{Rani2018} found that the emission above $20$ keV is dominated by a non-thermal component with possible 
jet origin. NGC 1275 also exhibits correlated variability 
in different wavebands \citep{Aleksic2014, Fukazawa2018}.
The broadband spectral energy distribution (SED) of NGC 1275 has been 
explained by one-zone synchrotron self Compton model 
\citep{ngc1275_fermi, Aleksic2014, Fukazawa2018, Tanada2018} 
or a structured jet \citep{tavecchio2014}.

In this work, we have carried out 
a long-term study of NGC 1275 
in $\gamma$-rays utilising $11$ years  
of {\sl Fermi}-LAT observations and 
identified different activity states. 
This source was also observed in X-rays, UV and/or optical band 
by {\sl Swift}-XRT and {\sl Swift}-UVOT multiple times during 
{\sl Fermi} observing period. 
India's first multi-wavelength astronomical 
observatory ``{\sl AstroSat}'' \citep{agrawal2006,singh2014,rao2016}, 
also observed the source twice. 
{\sl AstroSat} is 
capable of observing the sky  
simultaneously over a wide range of energies covering  
from near-UV (NUV) and far-UV (FUV) bands using 
the Ultra-Violet Imaging Telescope 
(UVIT; \citealt{kumar2012,tandon2017}), soft 
X-ray band with the Soft X-ray Telescope (SXT; \citealt{singh2017}) to 
the hard X-ray band with the Large Area X-ray Proportional Counter
(LAXPC; \citealt{yadav2016a,antia2017}) and Cadmium-Zinc-Telluride 
Imager (CZTI; \citealt{vadawale2015,rao2017}).

We have carried out construction and modelling 
of average broadband SEDs during different states 
as identified in the $\gamma$-ray light curve 
and during {\sl AstroSat} observing periods  
to understand the long term behaviour of this source.  
Details of the analysis of 
multi-wavelength data used in this work are given in Section 2. 
In Section 3, we discuss our findings, followed by a conclusion  
in Section 4. 

\section{DATA ANALYSIS AND RESULTS}\label{data_analysis}
\subsection{GeV DATA}\label{sec:data_fermi}

We used the $\gamma$-ray data from {\sl Fermi}-LAT \citep{Atwood2009} covering the 
period from 2008-08-04 to 2019-08-04 ($11$ years)  
in the energy range from $100$ MeV to $100$ GeV. 
The data reduction is performed
using the Fermitools version 1.2.23 
The current version of {\sl  Fermi}-LAT data, 
pass 8 P8R3, was used for analysis \citep{Bruel2018}. 
Instrument Response Function (IRF) \texttt{P8R3$\_$SOURCE$\_$V2}  
was used for source class event selection. 
A $15^{\circ} \times 15^{\circ}$ 
region of interest (ROI) 
centred at 
NGC 1275 was defined, and standard cuts were 
applied to select the good time intervals 
($z_{max} < 90^{\circ}$, \texttt{DATA$\_$QUAL$ > 0$} 
and \texttt{LAT$\_$CONFIG$==1$}).  
\texttt{Fermipy} version 0.19.0 \citep{fermipy} 
was used to calculate the light curve and spectra using 
binned likelihood analysis. A spatial binning of $0.1^{\circ}$ pixel$^{-1}$ 
and eight logarithmically-spaced energy bins per decade were chosen. 
Our initial model, generated using \texttt{make4fglxml.py}\footnote{\url{https://fermi.gsfc.nasa.gov/ssc/data/analysis/user/}},
consists of all $\gamma$-ray point sources 
within $20^{\circ}$ of the ROI centre included 
in the 4FGL-DR2 catalogue \citep{4FGL-DR2} and standard templates for Galactic 
diffuse emission model ({\sl gll\_iem\_v07.fits}) and isotropic diffuse emission ({\sl iso\_P8R3\_SOURCE\_V2\_v1.txt}) 
as used in the fourth {\sl Fermi} catalogue (4FGL: \citet{4FGL}).

\subsubsection{Eleven Years Averaged Spectrum} \label{avg_gamma_spec}

To calculate the average flux for the $11$ years dataset, 
we begin with an initial automatic optimisation of the ROI by 
iteratively fitting the sources using the \texttt{optimize} method of \texttt{Fermipy}. 
It is recommended by the \texttt{Fermipy} developers to run this method at the start 
of the analysis ``to ensure that all parameters are close to their global likelihood maxima'' 
\footnote{https://fermipy.readthedocs.io/en/latest/}.
After that, both normalisations and spectral parameters of the sources within $5^{\circ}$ 
and only normalisations of the sources lying within $12^{\circ}$ 
of the ROI centre were left to vary.
We freeze the spectral parameters for sources having 
Test Statistic (TS) $< 1$ or a predicted number of counts (Npred) 
after initial optimisation less than $10^{-3}$. 
Following \citet{Meyer2019}, the normalisations of the Galactic and isotropic diffuse backgrounds, 
including the spectral index of the Galactic diffuse 
background template, were left free during the fit.  
A TS map was generated using \texttt{findsource} tool 
of \texttt{Fermipy} to search for additional point sources, that are not 
present in the 4FGL-DR2 catalogue. No new sources were detected with 
TS $\geq 25$.

The source was modelled using simple power-law 
$$\frac{dF}{dE}=N \left(\frac{E}{E_{0}}\right)^{-\alpha}$$ 
and log-parabola
$$\frac{dF}{dE}=N \left(\frac{E}{E_{b}}\right)^{-\alpha-\beta \log{\left(\frac{E}{E_{b}}\right)}}$$ 
models.

Here, $\dfrac{dF}{dE}$ and $N$ are the differential flux and normalisation 
factor, respectively in the unit of photon cm$^{-2}$s$^{-1}$ MeV$^{-1}$. 
$E$ is the energy, $E_{0}$ and $E_{b}$ are the scale and break value, respectively in the unit of MeV.  
$\alpha$ and $\beta$ are the spectral parameters. 
The source was considered to be detected if its TS $> 25$, which corresponds 
to $\sim 4\sigma$ confidence \citep{mattox96}. The source spectrum is 
considered significantly curved if two times the difference in 
log-likelihood value for log-parabola and log-likelihood value for
power-law ($TS_{curve}$) is greater than $16$ \citep{acero20153fgl}.
The average spectrum of NGC 1275, utilising the 
$10$ years of {\sl Fermi} observations is reported 
to be significantly curved \citep{4FGL-DR2}. 
Utilising the $11$ years data set, we also noticed a significant curvature 
in the source spectrum.

\subsubsection{Temporal Analysis} \label{temporal}

\begin{figure*}
\centering
\includegraphics[height=15cm,width = 12.0cm]{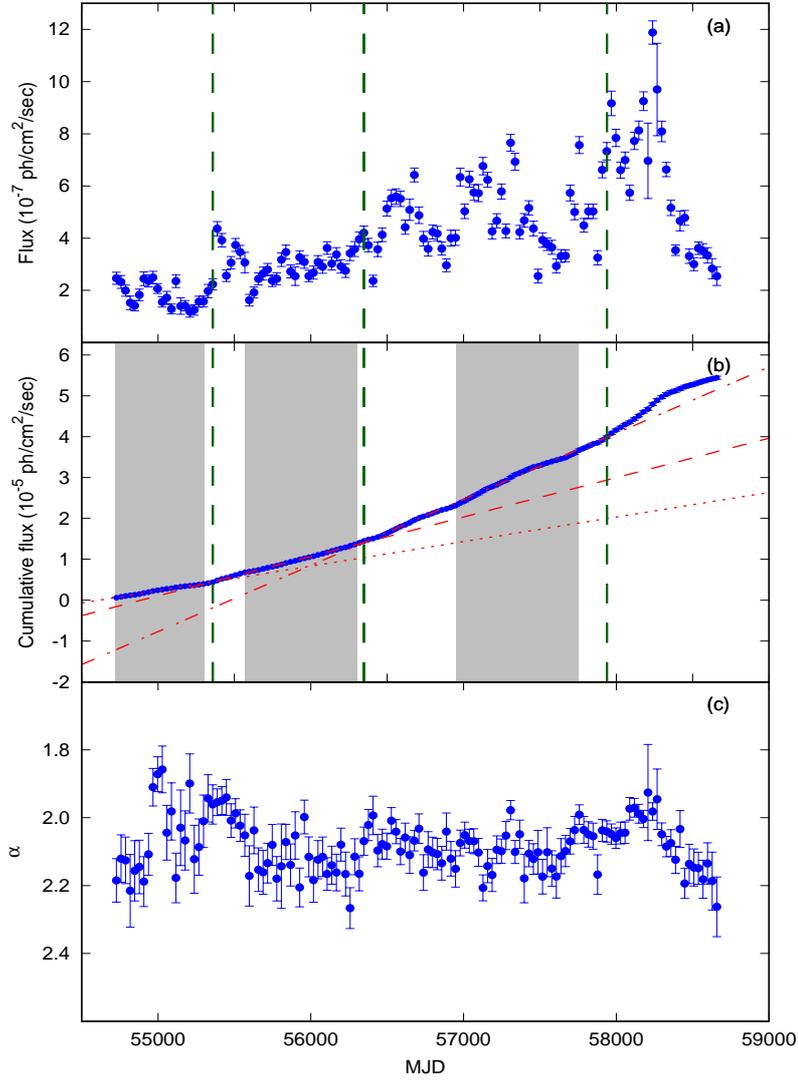}
\caption{(a): Monthly averaged $\gamma$-ray light curve ($100$ MeV-$100$ GeV) for NGC 1275 utilising 
$11$ years of {\sl  Fermi} observation (2008-2019). (b): Plot of cumulative flux with time. 
Data in shaded regions was considered for fitting linear function. 
The best fit linear functions in the three 
shaded regions are represented by the dotted line (slope: $6.0$ $\times 10^{-9}$ ph/cm$^2$/sec/day), 
	dashed line (slope: $9.6$ $\times 10^{-9}$ ph/cm$^2$/sec/day) 
and dash-dotted (slope: $1.6$ $\times 10^{-8}$ ph/cm$^2$/sec/day) line respectively. 
The vertical lines in (a) and (b) define the boundary of the different states.
(c) Variation of the monthly averaged spectral 
parameter $\alpha$ for NGC 1275 utilising $11$ years of {\sl  Fermi} observation (2008-2019).}
\label{ngc1275_monthly_states}
\end{figure*}

The monthly averaged $\gamma$-ray light curves in the energy bands $100$ MeV to 
$100$ GeV, $100$ MeV to 
$1$ GeV, and $1$ GeV to 
$100$ GeV were computed using \texttt{lightcurve} 
tool of \texttt{Fermipy}. The best fit model obtained for the $11$ years dataset 
considering a power-law spectrum for NGC 1275 was used as an input model.   
While constructing the 
$\gamma$-ray lightcurve in $100$ MeV to 
$100$ GeV energy band, both 
normalisations and spectral parameters of 
the sources within 
$3 ^{\circ}$ and only normalisations of 
sources within $3 ^{\circ}$ to $12 ^{\circ}$ from ROI centre 
were left free to vary in the input model of each time bin.  
For $\gamma$-ray lightcurves in $100$ MeV to 
$1$ GeV energy band and 
$1$ GeV to 
$100$ GeV energy band,  
only normalisations of the 
sources within 
$12 ^{\circ}$ from ROI centre 
were left free to vary in the input model for each time bin. 
For all the three light curves, 
the normalisations of the Galactic and isotropic 
diffuse emission models were left free, and the 
spectral index of the Galactic emission was frozen to the 
$11$ years averaged value. 
A signature of increase in the baseline flux was noticed 
from 2008 to 2017 (Fig~\ref{ngc1275_monthly_states}-a). 
This was followed by a large, broad flare until 2019.
 
In a previous work, \citet{Tanada2018} studied $\gamma$-ray variability 
of NGC 1275 utilising $\sim8$ years of {\sl Fermi} observations ($2008-2016$) and 
defined two epochs (epoch A: before February 2011 and epoch B: after February 2011) 
in the light curve based on the variations in the 
spectral indices. 
In this work, we have defined different activity states in the 
monthly averaged $\gamma$-ray light curve utilising $11$ years 
of {\sl Fermi} observations based on the increase in the baseline flux nature. 
To define different states, we calculated the cumulative flux 
of the monthly averaged $\gamma$-ray light curve. 
A constant slope 
in cumulative lightcurve corresponds to insignificant   
variation in flux. Whereas, an increase/decrease in slope 
indicates a rise/fall in baseline flux. 
Visual inspection of the cumulative flux distribution suggests 
the presence of four distinct regions/states. The first 
three states exhibit linear feature with increase in slopes, 
whereas, a significant non-linearity was noticed beyond that (state four). 
To constrain the boundaries of these states, data from shaded regions, as 
shown in Fig~\ref{ngc1275_monthly_states}-b, were fitted 
with linear functions. 
These regions were chosen in a way to avoid edge and/or transition effect. 
The mid-point of the monthly bin that contains the intersection of the 
fitted linear functions in the first and the 
second intervals is considered as the upper boundary of state 1 (S1). Similarly, 
the mid-point of the monthly bin that contains the intersection of 
the fitted linear functions in the second and the 
third intervals is considered as the upper boundary of state 2 (S2). A noticeable  
deviation from the fitted linear function in interval 3 is considered as the 
upper boundary of state 3 (S3). 
Boundaries of these states are indicated in 
Fig~\ref{ngc1275_monthly_states}-a. 
Beyond state 3, a significant non-linearity was noticed in 
the cumulative flux, first a sharp increase followed by a gradual decrease. 
This region represents the state 4 (S4). 
The details of the boundaries of these states are mentioned in Table~\ref{ngc_3state_table}. 
Fig~\ref{ngc1275_monthly_states}-c shows the variation of the monthly averaged 
spectral parameter $\alpha$ of this source. 
The epoch A and epoch B as defined in \citet{Tanada2018} overlap with S1-S2 and S2-S3, respectively.

\begin{table*}	
\centering
	\caption{$\gamma$-ray flux and spectral indices for different activity states}
       	\label{ngc_3state_table}
\vspace{0.5cm}
	\begin{tabular}{|c|c|c|c|c|c|c|} 
     		\hline
		Interval   & Start Date               &End Date                 & Flux ($\times10^{-7}$) & alpha & beta & TS$_{\mbox{curve}}$ \\
                           &  MJD                     & MJD                     & ph/cm$^{2}$/sec &  & &\\
                 (1)   & (2)  & (3)   & (4) & (5) & (6) & (7)\\
                 \hline
                State 1 (S1) &$2008$ Aug 05$~(54683)$     &$2010$ June 11$~(55358)$  & $1.72\pm0.07$ &  $2.02\pm0.02$ & $0.04\pm0.01$ & $19.3$ \\
		State 2 (S2) &$2010$ June 11$~(55358)$    &$2013$ Feb 25$~(56348)$   & $2.74\pm0.08$ &  $2.07\pm0.01$ & $0.060\pm0.007$ & $79.4$ \\
		State 3 (S3) &$2013$ Feb 25$~(56348)$     &$2017$ July 04$~(57938)$  & $4.44\pm0.06$ &  $2.069\pm0.008$ & $0.059\pm0.005$ & $212.9$ \\
		State 4 (S4) &$2017$ July 04$~(57938)$    &$2019$ Aug 05$~(58700)$   & $5.01\pm0.18$ &  $2.02\pm0.02$ & $0.07\pm0.01$ & $129.8$ \\
		AS1     &    $2017$ Jan 12$~(57765)$      &$2017$ Jan 14$~(57767)$   & $8.6\pm1.5$ &    $2.0\pm0.1$ & -       &  $3.4$         \\
                AS2     &    $2017$ Sep 26$~(58022)$      &$2017$ Sep 28$~(58024)$   & $5.4\pm1.6$ &    $2.1\pm0.2$ & -       &  $0.002$        \\

                  \hline         
      \end{tabular} 

\end{table*}

\subsubsection{Time-resolved Spectrum}
 $\gamma$-ray spectral analysis was carried out in different activity states 
to study the behaviour of the source. The average flux in the four activity states 
as defined in the $\gamma$-ray light curve was calculated following the methodology 
used for spectral analysis of the entire data set (Sec.~\ref{avg_gamma_spec}). 
To calculate the average flux in an activity state, the spectral index of the 
Galactic diffuse background template was kept frozen to the $11$ years averaged value. 
For all the four states, it was found that the log-parabola spectral model was 
strongly preferred over the power-law model. The best fit values of spectral 
parameters are given in Table~\ref{ngc_3state_table}. The spectral parameter 
$\alpha$ of the source in all four states lies in the range $2.0-2.1$. An 
increase of average $\gamma$-ray flux was noticed from S1 to S4.  
$\gamma$-ray SEDs were constructed 
for $8$ logarithmically-spaced energy bins per decade 
in $100$ MeV to $100$ GeV energy band using the \texttt{sed} tool of 
\texttt{Fermipy} for these four states. 
The spectral index in each energy bin 
was frozen to the power-law approximation (local index) to the shape of the 
global spectrum, while normalisation was left to vary.
 
{\sl AstroSat} observed NGC 1275 on 2017 January 12-13 under 
Target of Opportunity (ToO: hereafter referred to as AS1) which 
falls in the later phase of the S3 state. 
NGC 1275 was observed again with {\sl AstroSat} on 2017 September 26-27 
(referred to as AS2 in the rest of the paper) which falls at the beginning of the S4 state. 
The $\gamma$-ray analysis during AS1 and AS2 
was carried out following a similar methodology as 
used in S1-S4 states. However, due to limited 
photon statistics, a simple power-law was 
considered to model the source spectrum 
during these AS1 and AS2 states. 
The normalisations and spectral parameters of 
sources within $3 ^{\circ}$, only 
normalisations of sources within $3 ^{\circ}$ to $12 ^{\circ}$ 
from ROI centre and, normalisations of the 
Galactic and isotropic emission were left free to vary. 
We freeze the normalisations and spectral parameters for 
sources having TS$ < 1$ or Npred after initial optimisation less than 
$10^{-3}$.  
While constructing the $\gamma$-ray SED during AS1 and AS2,  
two energy bins per decade in the full energy band were considered. 
Due to low photon statistics, the data in the last two energy bands in the 
$\gamma$-ray SED was merged into a single energy band. 
The flux and spectral parameters during AS1 and AS2 
observing periods are given in Table~\ref{ngc_3state_table}.
The $\gamma$-ray flux during AS1 and AS2 (2 days of observation) 
is higher than the average flux during the longer duration states (a few years of observation).

\subsection{X-ray DATA}\label{sec:data_xray}
For X-rays, we used data from the {\sl Swift} 
X-ray Telescope \citep[{\sl Swift}-XRT;][]{burrows2005} that covers
the energy range for $0.3 - 10$ keV 
as well as SXT and LAXPC onboard {\sl AstroSat}. 
The energy ranges of 
SXT and LAXPC are 
$0.3 - 8$ keV and $3 - 80$ keV,  
respectively.

\subsubsection{{\sl Swift}-XRT}

Though there was no continuous monitoring of this 
source in the X-ray band, a significant number of pointed 
observations were carried out by {\sl Swift}-XRT. \citet{Fukazawa2018}
 analysed archival {\sl Swift}-XRT data of NGC 1275 from 2007-2015. 
They derived flux for each observation ID and calculated the 
light curve. In this work, we have calculated the light curve 
in a similar way utilising all the archival {\sl Swift}-XRT photon 
counting (PC) mode observations \footnote{\url{https://heasarc.gsfc.nasa.gov/cgi-bin/W3Browse/w3browse.pl}} 
covering S1, S2 and S3 states. Only one observation is present in S1, which 
is also studied by \citet{Fukazawa2018}. There are $16$, and $50$ 
PC mode observations present during S2 and S3 state, respectively. 
The analysis results of $14$ and $4$ PC mode observations during S2 and S3, respectively were given in \citet{Fukazawa2018}.  
During the AS1 state, there are no {\sl Swift}-XRT PC mode 
observations available. However, one {\sl Swift}-XRT observation in windowed timing (WT) mode 
(Obs ID: 00031770015, exposure time: $\sim239$ s) was present during this 
period, which was used in this work. 
We processed the XRT data using the \textsc{xrtdas} (v 3.4.1) package distributed 
under \textsc{heasoft} (v 6.24). The task \textsc{xrtpipeline} (v 0.13.4) was used to clean and 
calibrate level 1 data files with standard filtering criteria 
\footnote{\url{https://swift.gsfc.nasa.gov/analysis/xrt\_swguide\_v1\_2.pdf}} 
and using calibration files \textsc{caldb} (v 20190910). 
The source spectrum was binned to have $20$ counts per bin using \textsc{grppha}.
The XRT spectra were fitted in \textsc{xspec} version: 12.10.0c \citep{Arnaud1996}.

\begin{figure*}
\flushleft
\includegraphics[height=6cm,width = 8.5cm]{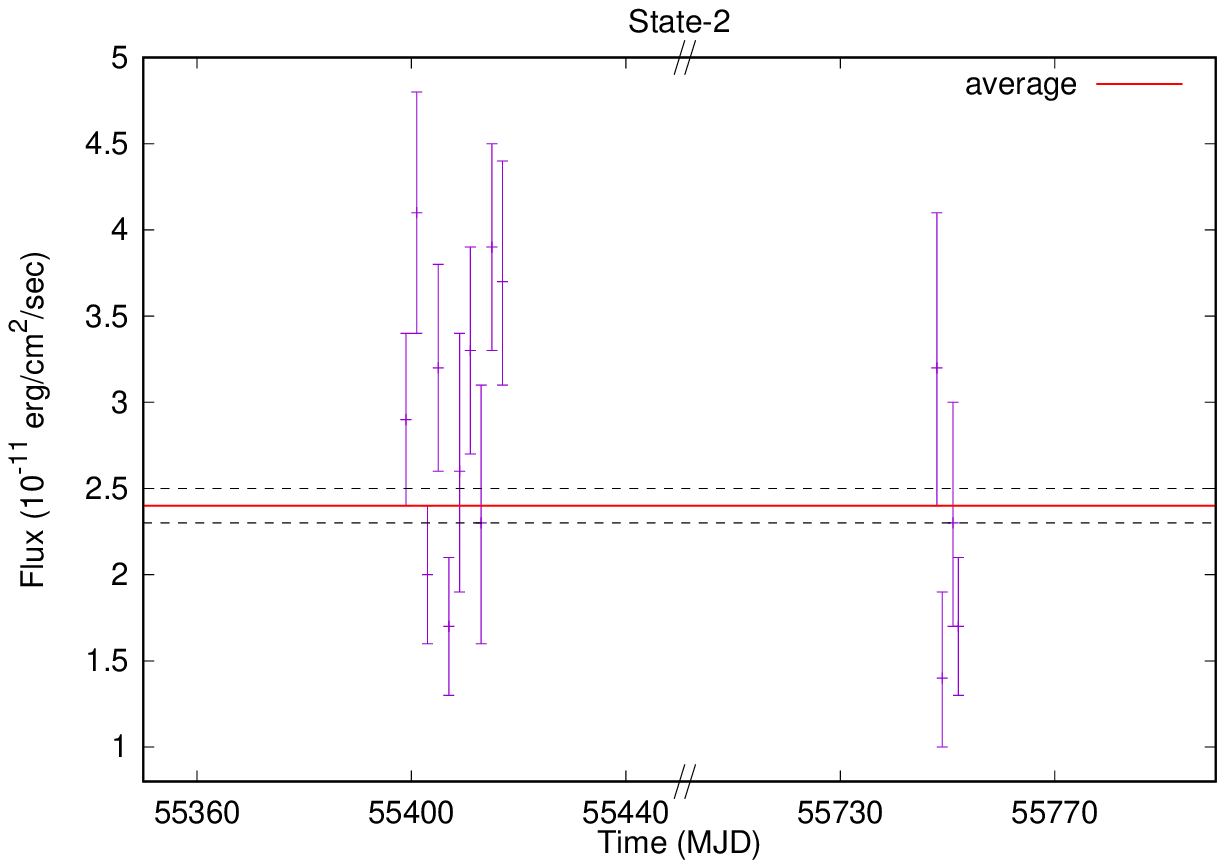}
\includegraphics[height=6cm,width = 8.5cm]{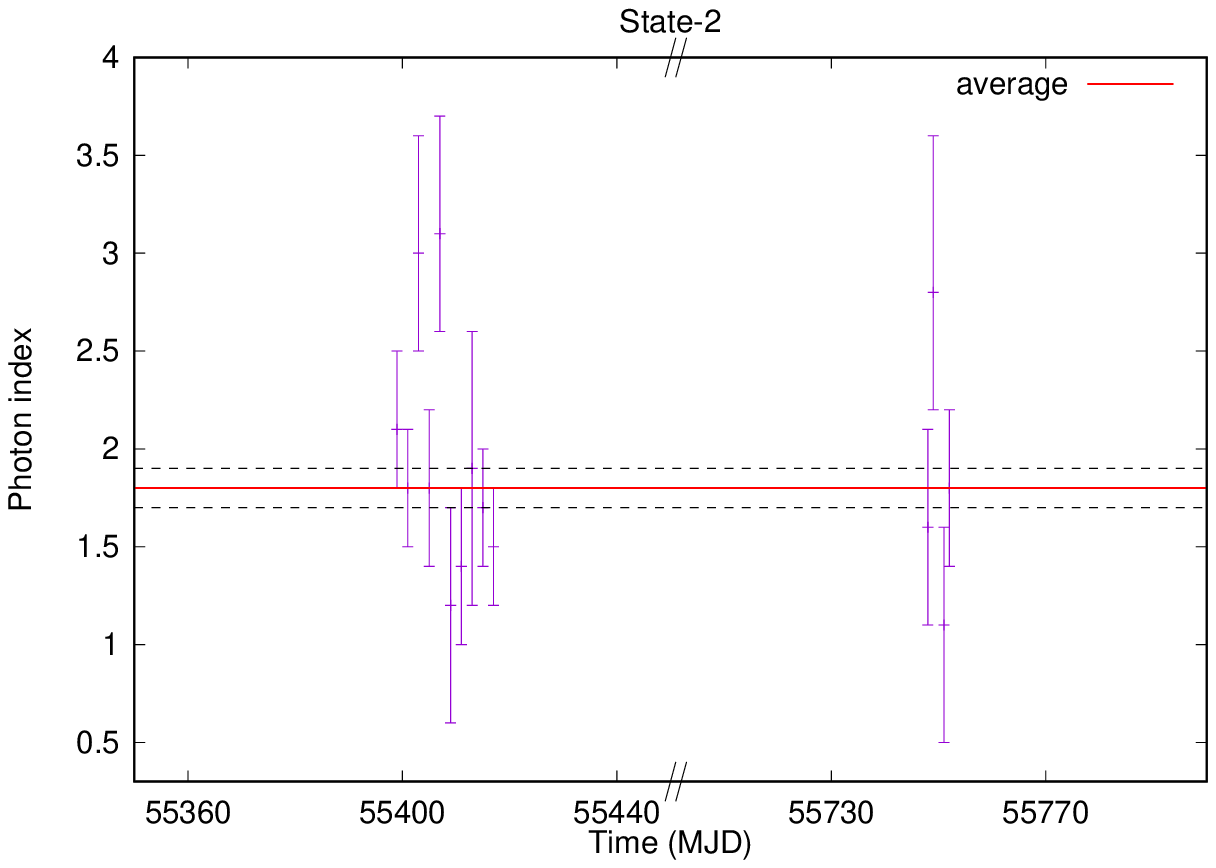}
\caption{{\sl Swift}-XRT light curve and variation of photon index for NGC 1275 for state 2. 
Left panel: Variation in $1.6-10$ keV flux. Right panel: Variation in photon index. 
The solid horizontal lines in both plots represent the weighted average and dashed horizontal lines 
represent the error in weighted average.}
\label{xrt-lc-s2}
\end{figure*}

\begin{figure*}
\flushleft
\includegraphics[height=6cm,width = 8.5cm]{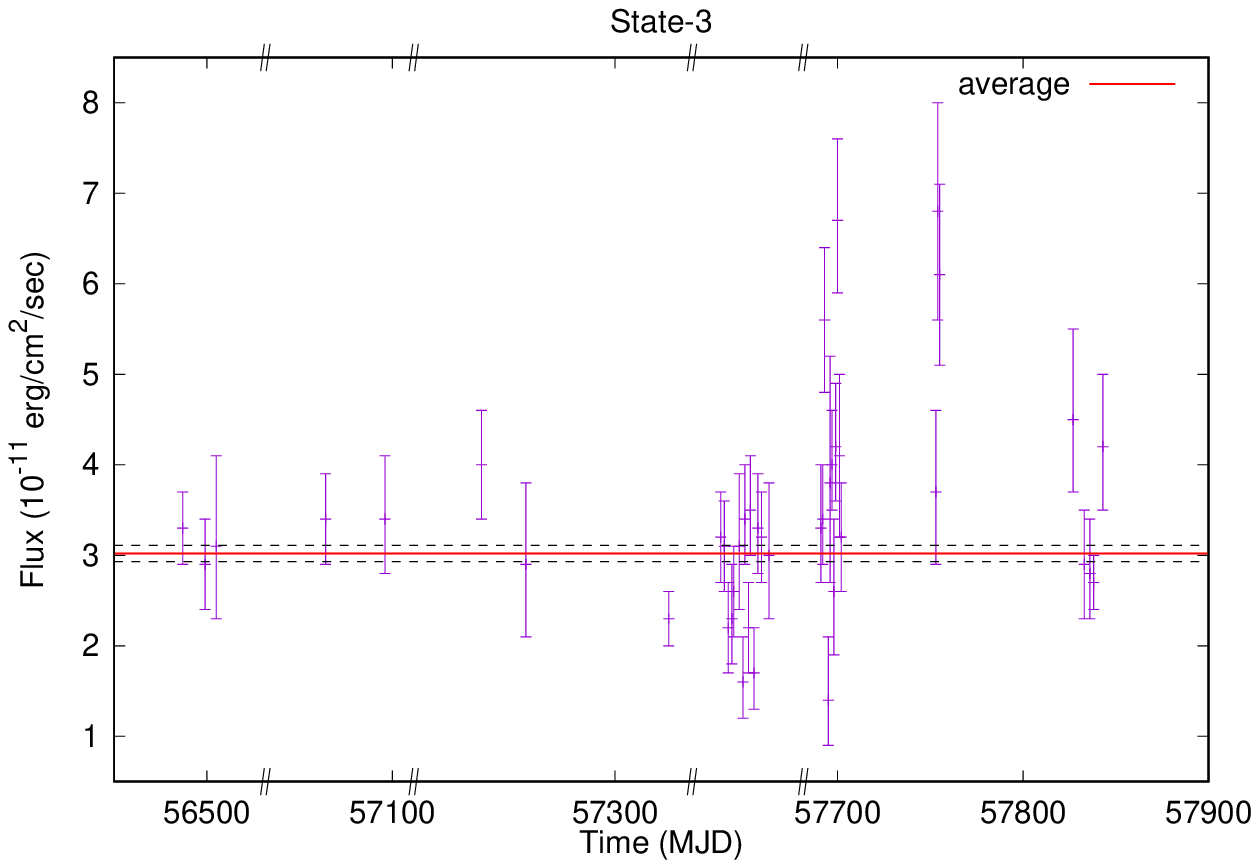}
\includegraphics[height=6cm,width = 8.5cm]{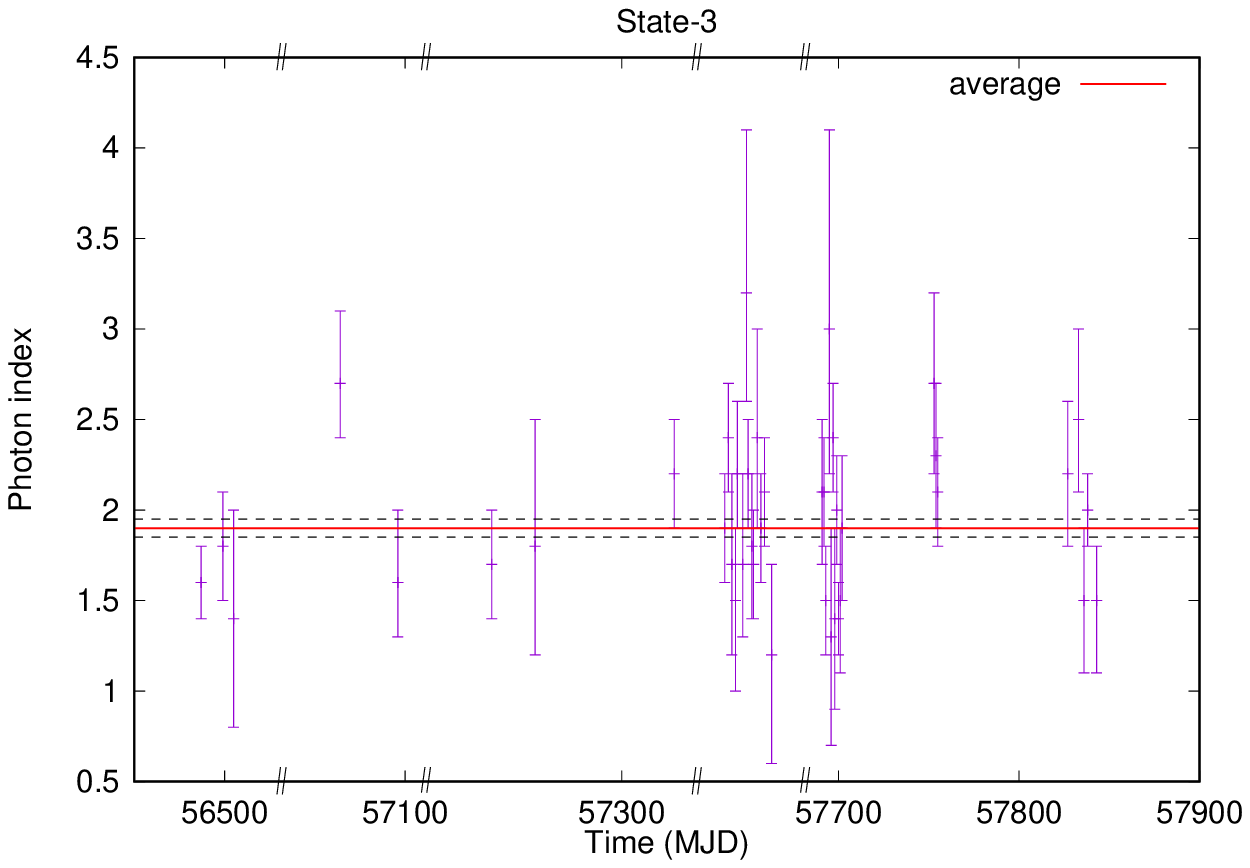}
\caption{{\sl Swift}-XRT light curve and variation of photon index for NGC 1275 for state 3. 
Left panel: Variation in $1.6-10$ keV flux. Right panel: Variation in photon index. 
The solid horizontal lines in both plots represent the weighted average and dashed horizontal lines 
represent the error in weighted average.}
\label{xrt-lc-s3}
\end{figure*}

For PC mode data, the source spectrum was derived considering an annular 
region from $12\arcsec - 26\arcsec$. The central $12\arcsec$ region is blocked since the 
PC mode data suffers from pile-up events. Background spectra were extracted 
in $60\arcsec - 65\arcsec$ from the centre to account for the cluster emission. Considering 
the significant difference in the response functions of PC mode and WT mode of 
{\sl Swift}-XRT below $1.6$ keV \citep{Godet2009}, following \citet{Fukazawa2018}, 
we restricted our analysis to $1.6 -10.0$ keV energy band. 
The XRT spectra were fitted with model ``phabs'', single temperature ``apec'' and 
power-law to account for Galactic absorption, the emission from hot gas in the 
Perseus cluster and the AGN emission, respectively. The Galactic hydrogen column density 
($N_H$) frozen to $1.5\times 10^{21}$ cm$^{-2}$ \citep{Yamazaki2013,Tanada2018}. We fix the 
temperature and abundance of the ``apec'' model to $4.1$ keV and $0.65$ solar, 
respectively \citep{Fukazawa2018} and estimate the normalisation of the ``apec'' model 
in each observation. The ``apec'' normalisation was then frozen to the median 
value of $0.0099$ \footnote{The ``apec'' normalisation, in units of $\frac{10^{-14}}{4\pi[D_A(1+z)]^2}\int n_e n_H dV$, 
where $D_A$ is the angular diameter distance to the source in cm, $n_e$ and $n_H$ are the electron and 
Hydrogen densities in cm$^{-3}$, respectively (followed throughout the paper).}
, and the AGN flux and photon index was derived in each observation ID. 
AGN flux was derived using ``cflux'' routine in \texttt{XSPEC}. 
No significant variation in photon indices was noticed during S2 and S3 states. 
The source flux exhibits noticeable variations in both S2 and S3 states with a 
fractional variance of $23$\%$\pm7$\% and $29$\%$\pm4$\%, respectively. 
Since our prime objective is to study the average source properties during these states, 
we used the flux and photon indices from each ID to calculate the average 
flux and photon indices during S2 and S3 states. 
The flux and photon index 
values in each of these individual observation IDs are summarised in Table~\ref{xrt_observ_pc}. 
Fig.~\ref{xrt-lc-s2} and Fig.~\ref{xrt-lc-s3} represents the flux and photon index values 
in each ID during S2 and S3 respectively, 
where the horizontal solid lines and dashed lines represent the weighted mean and error 
in weighted mean, respectively. The cases where power-law parameters could not be constrained 
are not considered while calculating the weighted mean and are also not shown 
in Table~\ref{xrt_observ_pc}, Fig.~\ref{xrt-lc-s2} and Fig.~\ref{xrt-lc-s3}. 
The values of average flux and photon indices are given 
in Table~\ref{table4}. 
Similar to $\gamma$-rays, a signature of average flux enhancement 
without any significant change in the photon index from S1 to S3 was observed.

The WT mode data during the AS1 period was analysed following 
the methodology described in \citet{Fukazawa2018}. 
The spectra was extracted within $0.3\arcmin$ of NGC 1275. 
The corresponding extraction region is $36\times1416$ arcsec$^2$ in the sky.  
All the PC mode data was used to extract the background spectra from the 
same sky region as that of WT mode spectra after excluding the central $36\times36$ arcsec$^2$, 
and subtracted from the WT mode spectra. 
A $3\%$ systematic error was considered during the spectral analysis 
of WT mode data.
Parameters of the ``apec'' model 
namely, temperature, abundance, and normalisation 
were fixed to $4.0$ keV, $0.60$ solar and $0.0208$, respectively \citep{Fukazawa2018}. 
The best fit values of flux and photon index are given in Table~\ref{table4}.
Recently, \citet{Imazato2020} have presented {\sl Swift}-XRT light 
curves of NGC 1275. From visual inspection of their light curves, 
for the data in common, we found good
agreement between our results and that of \citet{Imazato2020}.

\subsubsection{{\sl AstroSat}}
The Level 1 SXT data was analysed 
using \textsc{``sxtpipeline''} of the SXT 
software \textsc{``as1sxtlevel2-1.4b''}.
The clean events were merged 
using the \textsc{``sxtpyjuliamerger\_v01''}. 
A circular region of $15'$ radius was used as a source region, and 
for the background, the instrument team provided {\sl ``SkyBkg\_comb\_EL3p5\_Cl\_Rd16p0\_v01.pha''} file was used.
The ancillary response file was 
created by using \textsc{``sxteefmodule\_v02''}. 
A $2\%$ systematic error was considered during spectral analysis.

LAXPC data was analysed using the analysis software 
\textsc{``laxpc\_soft''} package (May 19, 2018 version) available at the {\sl AstroSat} 
Science Support Cell \footnote{\url{http://astrosat-ssc.iucaa.in/?q=data\_and\_analysis}}. 
Standard procedures were used to reduce the Level 1 
data \citep{yadav2016b,antia2017}. The spectra for the source 
and background in the energy range of $4.0-13.0$ keV were created using the layer 1 
data of PCU unit 20.
The energy resolution of LAXPC is $15\%$. 
Therefore, it is recommended by the LAXPC instrument team 
that the spectrum should be re-binned such that the energy bin 
width is at least $5\%$ of the central energy 
(energy grouping factor $0.05$). Hence, 
in our analysis, we considered a value of $0.05$ for energy grouping factor.  
Since the size of the emission region is large, 
the systematic error of LAXPC 
is expected to be high. We considered a $3\%$ systematic 
error for LAXPC spectral analysis. 

Due to the large source extraction region of 
{\sl AstroSat}-SXT ($\sim 15'$ radius) and 
{\sl AstroSat}-LAXPC ($1^{\circ} \times 1^{\circ}$ field of view), 
the observed X-ray flux of NGC 1275 suffers significant contamination 
from the Perseus cluster.
Therefore, to obtain intrinsic emission 
from the nucleus of NGC 1275, 
it is essential to constrain cluster parameters. 
For SXT and LAXPC analysis, the 
cluster abundance was frozen to the value of $0.42$ solar, which was obtained by averaging the data 
from Fig.~6 of \citet{Churazov2003} over a region of $15\arcmin$ radius. This value of abundance is consistent 
with other studies \citep{Schmidt2002,Nishino2010}.  
The temperature and abundance of the ``apec'' model, 
were constrained utilising simultaneous {\sl Swift}-XRT observation. 

As during {\sl Swift}-XRT analysis data below $1.6$ 
keV was not considered; we restricted our SXT data 
analysis to $1.6-7.0$ keV energy band. For SXT analysis 
during the AS1 observing period (exposure time $\sim 33$ ks), 
the photon index and the unabsorbed flux of the power-law component  
were kept frozen to the values derived from the simultaneous {\sl Swift}-XRT observation to estimate the 
temperature ($5.60\pm0.09$ keV) and normalisation ($0.708\pm0.003$) of the ``apec'' model.  
For SXT analysis during the AS2 observing period (exposure time $\sim 17$ ks), 
the temperature and normalisation were kept frozen to the 
best fit values obtained during the AS1 observing period, and the unabsorbed flux from NGC 1275 
was obtained (Table~\ref{astrosat_x_ray_results}). 

During LAXPC analysis in the AS1 period (exposure time $\sim 50$ ks), 
the photon index and the unabsorbed flux of the power-law component 
(emission from NGC 1275) were kept frozen to the values derived from simultaneous 
{\sl Swift}-XRT observation in $4.0-13.0$ keV energy range and the best fit values of the temperature 
($5.9\pm0.1$ keV) and normalisation ($0.75\pm0.02$) of the ``apec'' model were obtained.  
During the AS2 observing period (exposure time $\sim 16$ ks), the ``apec'' temperature and normalisation were 
kept frozen to the values obtained during the AS1 period. Due to low photon statistics, 
the photon index of the source could not be constrained with LAXPC observation. Therefore, 
the power-law photon index was kept frozen to the best fit value obtained during simultaneous SXT observation. 
The unabsorbed flux from NGC 1275 in $4.0-13.0$ keV energy 
band is given in Table~\ref{astrosat_x_ray_results}.

{\footnotesize
\clearpage
\onecolumn
\begin{longtable}{l c c c c r}
\caption{Summary of {\sl Swift}-XRT PC mode observations \label{xrt_observ_pc}. $\star$ marks the observation IDs analysed by \citet{Fukazawa2018}.}\\
\hline
State	& Sequence No. & Date	& Exposure Time & Flux (in $\times 10^{-11}$ erg/cm$^2$/sec)	& Photon index \\
 &              &        &     (seconds) &      & \\
\hline
S1	&	$00030354003^*$	     &  2009-12-30 (55195)	 & 4328               & $1.7\pm0.4$ &$1.9\pm{0.5}$ \\
&              &        &     &      & \\
\hline
S2	&	$00031770001^*$	     &  2010-07-22 (55399)       & 2198	    	& $2.9\pm0.5$         & $2.1_{-0.3}^{+0.4}$\\
	&	$00031770002^*$      &  2010-07-24 (55401)       & 2048     	& $4.1\pm0.7$         & $1.8\pm0.3$\\
	&	$00031770003^*$      &  2010-07-26 (55403)       & 2045     	& $2.0\pm0.4$         & $3.0_{-0.5}^{+0.6}$\\
	&	$00031770004^*$      &  2010-07-28 (55405)       & 2183     	& $3.2\pm0.6$         & $1.8\pm0.4$\\
	&	$00031770005$        &  2010-07-30 (55407)       & 2125     	& $1.7\pm0.4$         & $3.1_{-0.5}^{+0.6}$\\
	&	$00031770006^*$      &  2010-08-01 (55409)       & 2119     	& $2.6_{-0.7}^{+0.8}$ & $1.2_{-0.6}^{+0.5}$\\
	&	$00031770007^*$      &  2010-08-03 (55411)       & 2412     	& $3.3\pm0.6$         & $1.4\pm0.4$\\
	&	$00031770008^*$      &  2010-08-05 (55413)       & 1998     	& $2.3_{-0.7}^{+0.8}$ & $1.9\pm0.7$\\
	&	$00031770009^*$      &  2010-08-07 (55415)       & 2113     	& $3.9\pm0.6$         & $1.7\pm0.3$\\
	&	$00031770010^*$      &  2010-08-09 (55417)       & 2118     	& $3.7_{-0.6}^{+0.7}$ & $1.5\pm0.3$\\
	&	$00091128002^*$      &  2011-07-06 (55748)       & 1349         & $3.2_{-0.8}^{+0.9}$ & $1.6\pm0.5$\\
	&	$00091128003^*$      &  2011-07-07 (55749)       & 1771     	& $1.4_{-0.4}^{+0.5}$ & $2.8_{-0.6}^{+0.8}$\\
	&	$00091128004^*$      &  2011-07-09 (55751)       & 3479     	& $2.3_{-0.6}^{+0.7}$ & $1.1_{-0.6}^{+0.5}$\\
	&	$00091128005^*$      &  2011-07-10 (55752)       & 4925     	& $1.7\pm0.4$         & $1.8\pm{0.4}$\\	
&              &        &     &      & \\

\hline  
S3	&	$00049799004^*$	   &	2013-07-14  (56487)  & 5282     &$3.3\pm0.4$           &  $1.6\pm0.2$\\
        &	$00049799005^*$    &	2013-07-26  (56499)  & 3077     &$2.9\pm0.5$   	    &  $1.8\pm0.3$\\
        &	$00049799006^*$    &	2013-08-01  (56505)  & 1573     &$3.1_{-0.8}^{+1.0}$   &  $1.4\pm0.6$\\
        &	$00092034001$      &	2015-02-11  (57064)  & 2003     &$3.4\pm0.5$           &  $2.7_{-0.3}^{+0.4}$\\
        &	$00092034002^*$    &	2015-03-15  (57096)  & 2150     &$3.4_{-0.6}^{+0.7}$   &  $1.6_{-0.3}^{+0.4}$\\
        &	$00092034003$      &	2015-07-25  (57228)  & 2005     &$4.0\pm0.6$           &  $1.7\pm0.3$\\
        &	$00092034004$      &	2015-08-18  (57252)  & 2008     &$2.9_{-0.8}^{+0.9}$   &  $1.8_{-0.6}^{+0.7}$\\
        &	$00081530001$      &	2015-11-03  (57329)  & 6428     &$2.3\pm0.3$           &  $2.2\pm0.3$\\
        &	$00034380001$      &	2016-02-19  (57437)  & 2475     &$3.2\pm0.5$           &  $1.9\pm0.3$\\
        &	$00034380002$      &	2016-02-21  (57439)  & 2480     &$3.1\pm0.5$           &  $2.4\pm0.3$\\
        &	$00034380004$      &	2016-02-23  (57441)  & 2417     &$2.2\pm0.5$           &  $1.7\pm0.5$\\
        &	$00034380005$      &	2016-02-25  (57443)  & 2757     &$2.3_{-0.5}^{+0.6}$   &  $1.5_{-0.5}^{+0.4}$\\
        &	$00034380006$      &	2016-02-26  (57444)  & 2780     &$2.6\pm0.5$           &  $2.2_{-0.3}^{+0.4}$\\
        &	$00034380007$      &	2016-02-29  (57447)  & 1703     &$3.1_{-0.7}^{+0.8}$   &  $1.7_{-0.4}^{+0.5}$\\
        &	$00034380008$      &	2016-03-02  (57449)  & 2382     &$1.6_{-0.4}^{+0.5}$   &  $3.2_{-0.6}^{+0.9}$\\
        &	$00034380010$      &	2016-03-03  (57450)  & 2030     &$3.4_{-0.5}^{+0.6}$   &  $2.2\pm0.3$\\
        &	$00034404001$      &	2016-03-05  (57452)  & 3976     &$2.2\pm0.5$           &  $1.8\pm0.4$\\
        &	$00034380012$      &	2016-03-06  (57453)  & 2362     &$3.5_{- 0.5}^{+0.6}$  &  $1.7\pm0.3$\\
        &	$00034380013$      &	2016-03-08  (57455)  & 2914     &$1.7_{-0.4}^{+0.5}$   &  $2.4_{-0.5}^{+0.6}$\\
        &	$00034380014$      &	2016-03-10  (57457)  & 2240     &$3.3_{-0.5}^{+0.6}$   &  $1.9\pm0.3$\\
        &	$00034380015$      &	2016-03-12  (57459)  & 2285     &$3.2\pm0.5$           &  $2.1\pm0.3$\\
        &	$00034404003$      &	2016-03-16  (57463)  & 2185     &$3.0_{-0.7}^{+0.8}$   &  $1.2_{-0.6}^{+0.5}$\\
        &	$00034765001$      &	2016-10-30  (57691)  & 1973     &$3.3_{-0.6}^{+0.7}$   &  $2.1\pm0.4$\\
        &	$00034765002$      &	2016-10-31  (57692)  & 1983     &$3.4_{-0.5}^{+0.6}$   &  $2.1\pm0.3$\\
        &	$00034765003$      &	2016-11-01  (57693)  & 1878     &$5.6\pm0.8$           &  $1.5\pm0.3$\\
        &	$00034765005$      &	2016-11-03  (57695)  & 1611     &$1.4_{-0.5}^{+0.7}$   &  $3.0_{-0.8}^{+1.1}$\\
        &	$00034765006$      &	2016-11-04  (57696)  & 1696     &$3.8_{-1.1}^{+1.4}$   &  $1.3\pm0.6$\\
        &	$00034765007$      &	2016-11-05  (57697)  & 1808     &$4.0_{-0.5}^{+0.6}$   &  $2.4\pm0.3$\\
        &	$00034765008$      &	2016-11-06  (57698)  & 2010     &$2.6_{-0.7}^{+0.8}$   &  $1.4\pm0.5$\\                   	  
        &       $00034765009$      &    2016-11-07  (57699)  & 1958     &$4.2_{-0.6}^{+0.7}$   &  $2.0\pm0.3$\\                
        &	$00034765010$      &	2016-11-08  (57700)  & 1613     &$6.7_{-0.8}^{+0.9}$   &  $1.4\pm0.2$\\
        &	$00034765011$      &	2016-11-09  (57701)  & 1543     &$4.1\pm0.9$           &  $1.5\pm0.4$\\
        &	$00034765012$      &	2016-11-10  (57702)  & 1935     &$3.2\pm0.6$           &  $1.9\pm0.4$\\
        &	$00087312001$      &	2016-12-31  (57753)  & 956     &$3.7_{-0.8}^{+0.9}$   &  $2.7\pm0.5$\\
        &	$00087311001$      &	2017-01-01  (57754)  & 634     &$6.8\pm1.2$           &  $2.3\pm0.4$\\
        &	$00087311002$      &	2017-01-02  (57755)  & 797     &$6.1\pm1.0$           &  $2.1\pm0.3$\\
        &	$00087311003$      &	2017-03-15  (57827)  & 1079     &$4.5_{-0.8}^{+1.0}$   &  $2.2\pm0.4$\\
        &	$00087312002$      &	2017-03-21  (57833)  & 1466     &$2.9\pm0.6$           &  $2.5_{-0.4}^{+0.5}$\\
        &	$00087311005$      &	2017-03-24  (57836)  & 2420     &$2.8_{-0.5}^{+0.6}$   &  $1.5\pm0.4$\\
        &	$00087312004$      &	2017-03-26  (57838)  & 4797     &$2.7\pm0.3$           &  $2.0\pm0.2$\\
        &	$00087312005$      &	2017-03-31  (57843)  & 2143     &$4.2_{-0.7}^{+0.8}$   &  $1.5_{-0.4}^{+0.3}$\\
&              &        &     &      & \\
 \hline                                              

\end{longtable}
}

\clearpage
\twocolumn

\begin{table}
\centering
\caption{{\sl Swift}-XRT analysis results for different activity states}
\label{table4}
\begin{tabular}{|l|c|r|} 
\hline
Activity state          &F$_{\mbox{1.6-10.0 keV}}$$^{a}$    &$\Gamma_{\mbox{1.6-10.0 keV}}$$^{b}$           \\
                  
\hline                                                                                                                               
State 1	           & $1.7\pm0.4$    	      &$1.9\pm0.5$      	         \\                                     
                                                                                                              
State 2            &$2.4\pm0.1$               &$1.8\pm0.1$                           \\

State 3            &$3.02\pm0.09$    	      &$1.90\pm0.05$      	             \\

AS1		   &$1.8_{-0.7}^{+0.9}$       &$1.9_{-0.8}^{+0.9}$      	       \\
                                                                                                              
\hline
\end{tabular}

$^{a}${ 1.6-10.0 keV unabsorbed flux in units of $10^{-11}$ erg cm$^{-2}$s$^{-1}$}\\
	$^{b}${ 1.6-10.0 keV photon index of power-law model}

\end{table}

\begin{table}
\centering
\caption{{\sl AstroSat}-SXT and LAXPC analysis results during AS2 period}
\label{astrosat_x_ray_results}
\begin{tabular}{lcccr} 
\hline
Instrument       &Energy range &F$^{a}$         &$\Gamma^{b}$      &$\chi^{2}$/dof        \\
                  & (keV)       &                &                  &                      \\
\hline
SXT          & $1.6-7.0$   &$3.0\pm0.3$   &$2.7\pm0.3$           &$496.89$/$439$     \\
           
LAXPC        & $4.0-13.0$  &$0.6\pm0.4$  &$2.7^{c}$            &$9.64$/$13$     \\

\hline
\end{tabular}
\\
$^{a}${Unabsorbed flux in units of $10^{-11}$ erg cm$^{-2}$s$^{-1}$}\\
$^{b}${Photon index of power-law model}\\
$^{c}${Photon index kept frozen to the best fit value obtained during simultaneous SXT observation.}\\

\end{table}

\subsection{UV/Optical data}\label{sec:data_optical}

\begin{table}
\centering
\caption{{\sl Swift}-UVOT flux values in different UVOT filters for different activity states}
\label{uvot_results}
\begin{tabular}{ccccc} 
\hline
Filters & S1$^{a}$               & S2$^{a}$                       & S3$^{a}$        & AS1$^{a}$\\
\hline

V    & $-$		& 	   $10.5\pm0.3$   & 	   $  13.2\pm0.4$  &    $-$	\\
B    & $-$              &          $7.4\pm0.3 $   &        $  9.1\pm0.3$  &    $-$	\\
U    & $-$              &          $4.3\pm0.2 $   &        $  6.2\pm0.2$  &    $-$	\\
UVW1 & $1.9\pm0.1$      &          $2.9\pm0.2 $   &          $  3.9\pm0.2$  &     $-$ \\
UVM2 & $-$              &          $3.0\pm0.1 $   &        $  3.6\pm0.2$  &    $-$ \\
UVW2 & $-$              &          $2.3\pm0.1 $   &        $  2.9\pm0.2$  &    $3.4\pm0.2$\\

\hline
\end{tabular}
\\
$^{a}${Flux in units of $10^{-26}$ erg cm$^{-2}$s$^{-1}$Hz$^{-1}$}\\
\end{table}

\begin{table}
\centering
\caption{{\sl AstroSat}-UVIT analysis results}
\label{uvit_result}
\begin{tabular}{lcr} 
\hline
Filter         &$\lambda_{mean}$(\AA)    &Flux$^{a}$ \\
\hline
CaF2-1 (F1)       &$1481$           &   $2.21\pm0.02$      \\           
Silica (F5)       &$1717$           &   $3.06\pm0.04$    \\
NUVB15 (F2)       &$2196$           &   $4.83\pm0.05$   	\\
NUVB4 (F5)        &$2632$           &   $4.32\pm0.03$   	\\                                                            

\hline
\end{tabular}
\\
$^{a}$Flux in $10^{-26}$ erg cm$^{-2}$s$^{-1}$Hz$^{-1}$\\

\end{table}

For data in the UV/optical bands, we used both {\sl Swift}-UVOT \citep{Roming2005} and UVIT onboard 
{\sl AstroSat}. From {\sl Swift}-UVOT, we have observations in $U$, 
$B$, $V$, $UVW1$, $UVW2$, and $UVM2$ filters for different activity states. 
The level 2 products \footnote{\url{https://heasarc.gsfc.nasa.gov/cgi-bin/W3Browse/w3browse.pl}} 
were analysed using different tasks, which are a part of \textsc{heasoft} (v 6.24), 
and 20170922 version of \textsc{caldb}.
\textsc{uvotimsum} task was used to merge the different observations during a particular
epoch. 

For photometry, a source region with radius $\sim$ point spread function 
was chosen to reduce cluster/host galaxy contribution. 
A circular source region with 
$3''$ radius centred at the source position 
and an annular background  
region with inner and outer radii of $15\arcsec$ and $20\arcsec$ was used.  
\textsc{uvotsource} task was used to get the 
source magnitude.   
The Galactic extinction was calculated using \cite{cardelli1989} 
and \cite{schlafly2011}.  The extinction corrected $AB$ magnitudes 
were then converted to flux (erg cm$^{-2}$ s$^{-1}$). 
During S2 and S3 states, the source was observed in all the 
{\sl Swift}-UVOT filters. In the state S2, the exposure times for 
$V$, $B$, $U$, $UVW1$, $UVM2$ and $UVW2$ filters are $\sim3$ ks, $\sim3$ ks, $\sim6$ ks, 
$\sim11$ ks, $\sim8$ ks and $\sim17$ ks, respectively. For the state 3, 
the exposure times for $V$, $B$, $U$, $UVW1$, $UVM2$ and $UVW2$ filters 
are $\sim5$ ks, $\sim5$ ks, $\sim23$ ks, $\sim33$ ks, $\sim23$ ks, $\sim30$ ks seconds respectively. 
However, during S1 and AS1 states, the source was only observed in 
$UVW1$ (exposure time: $\sim4$ ks) and $UVW2$ filters (exposure time: $232$ seconds), 
respectively. 
The Galactic extinction corrected fluxes in different 
UVOT filters during S1, S2, S3 and AS1 observing periods 
are given in Table~\ref{uvot_results}. 
{\sl Swift}-UVOT light curves of NGC 1275 were recently 
presented by \citet{Imazato2020}. As they have corrected for the
host galaxy contribution to the observed UV/optical emission, 
it is likely that the flux values quoted in Table~\ref{uvot_results} are
marginally brighter than that of \citet{Imazato2020}.

Similar to the X-ray and $\gamma$-ray bands, 
an increase in the average source flux was noticed from S1 to S3. 
Utilising the fluxes obtained in three optical and three UV filters and 
considering a power-law spectral shape, averaged energy spectral indices 
of optical and UV fluxes were derived in S2 and S3 states. An energy 
spectral index of $1.9\pm0.3$ and $1.65\pm0.01$ was obtained in the 
optical band for the S2 and S3 states respectively. Similarly, in the 
UV band, energy spectral index of $1.1\pm0.8$ and $1.2\pm0.4$ was 
obtained for the S2 and S3 states respectively. Hence, no appreciable 
change in spectral indices was noticed in these states.

NGC 1275 was observed with {\sl AstroSat}-UVIT under
AS2 observation in FUV filters: CaF2-1 (exposure time $\sim 4$ ks) 
and Silica (exposure time $\sim 8$ ks) and NUV filters: NUVB15 
(exposure time $\sim 8$ ks) and NUVB4 (exposure time $\sim 4$ ks). 
The science ready visual aspect corrected Level-2 images provided by Indian Space 
Science Data Centre (ISSDC) (processed by the pipeline version $6.3$) 
was used to carry out standard 
photometry using \textsc{iraf} \footnote{IRAF is distributed 
by the National Optical Astronomy Observatory, which is 
operated by the Association of Universities for Research 
in Astronomy (AURA)under a cooperative agreement with the National 
Science Foundation}. A circular aperture of $5$ pixels ($\sim2.0\arcsec$)  
and background region of $15-20$ pixels ($\sim6.0\arcsec - \sim 8.0\arcsec$) was used 
for photometry. The derived 
magnitudes were converted into fluxes \citep{tandon2017}
and corrected for Galactic extinction. 
The estimated flux was corrected for the chosen aperture size using Table~11 of \citet{Tandon2020}.
The Galactic extinction corrected fluxes 
in four UVIT filters during AS2 observing period 
are given in Table~\ref{uvit_result}.

\begin{figure*}
\includegraphics[height=6cm,width=8cm]{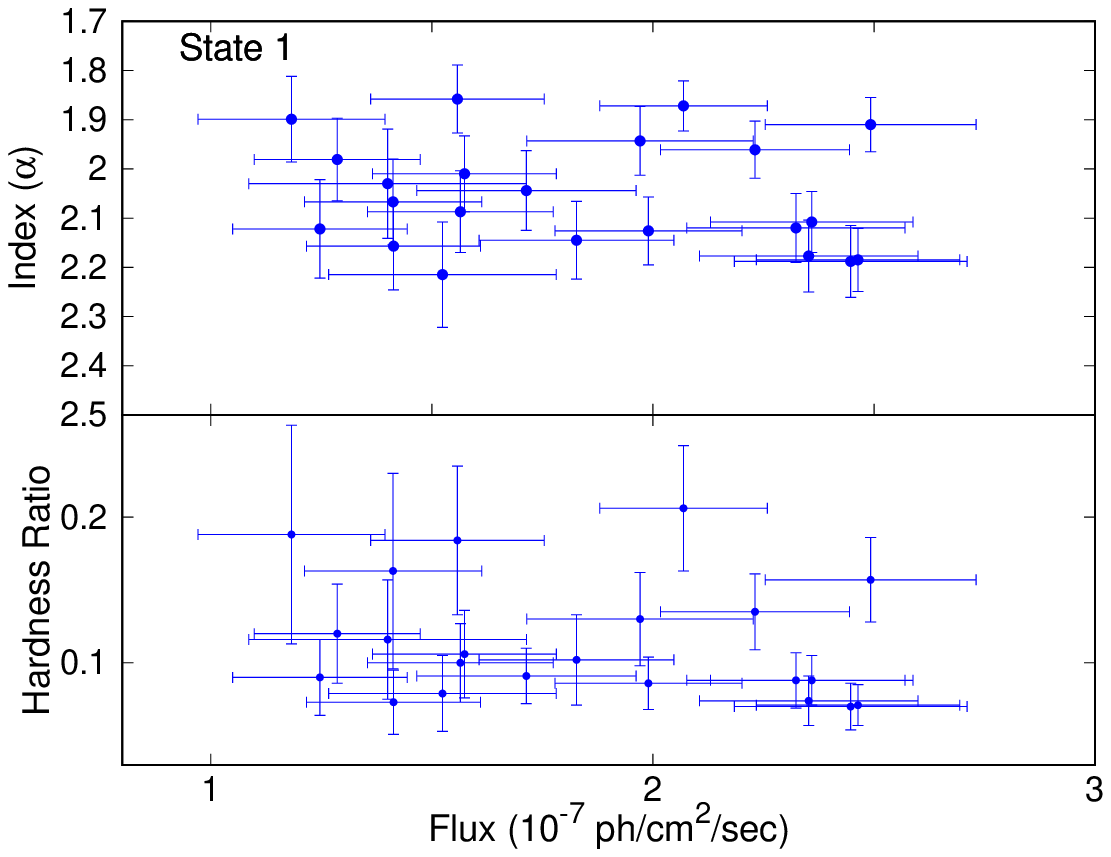}
\includegraphics[height=6cm,width=8cm]{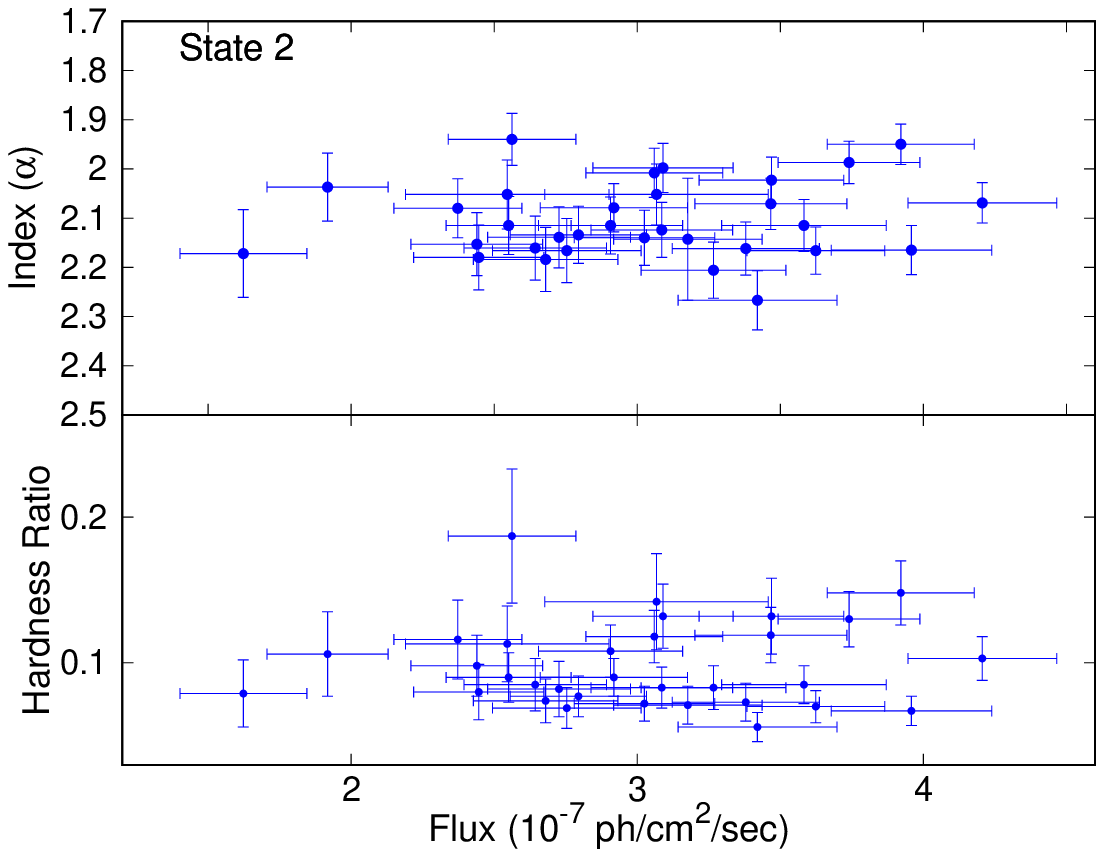} \\
\includegraphics[height=6cm,width=8cm]{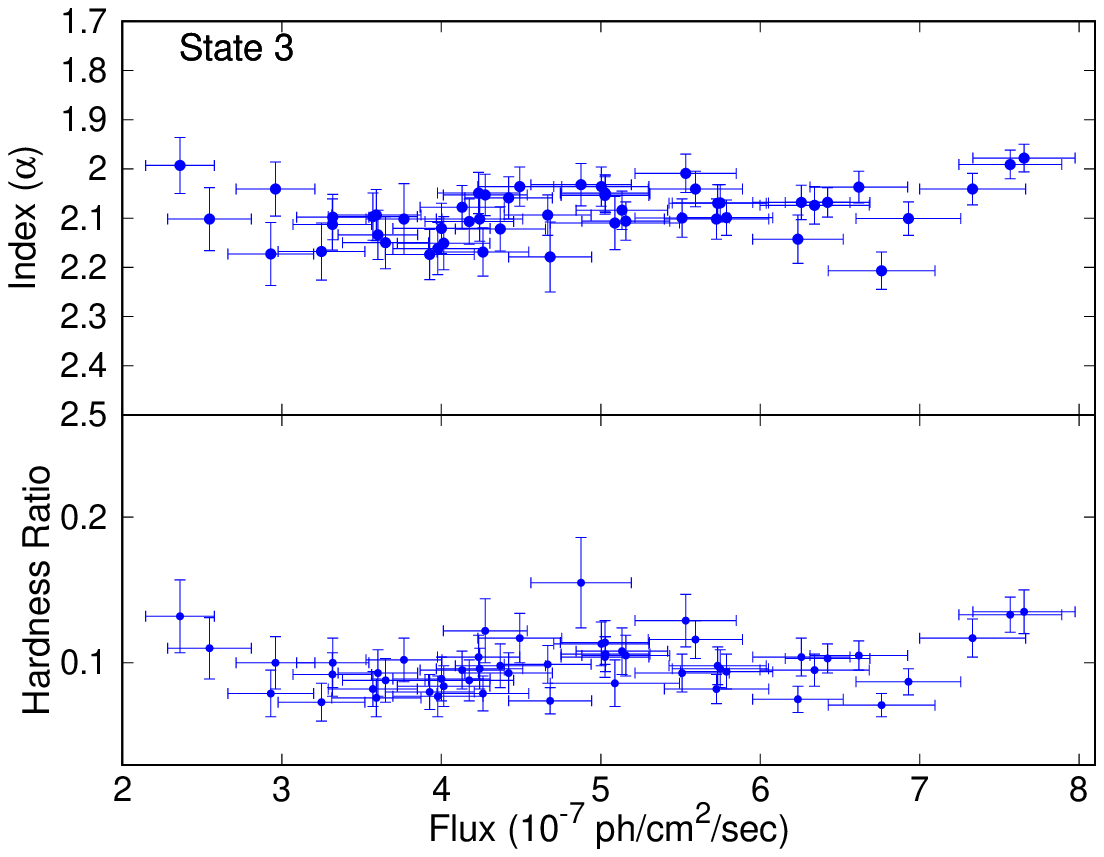}
\includegraphics[height=6cm,width=8cm]{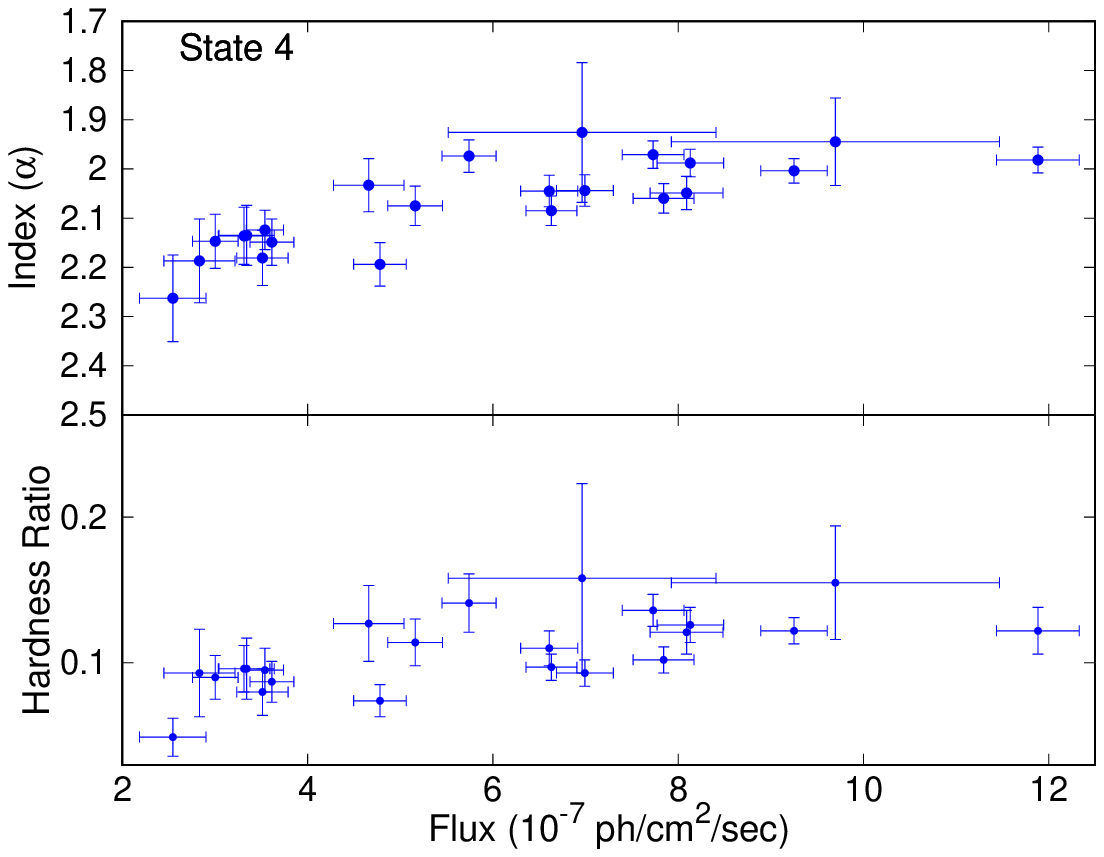} \\
\caption{Variation of monthly averaged $\gamma$-ray flux with photon index (top panel in each sub-figure) and 
monthly averaged $\gamma$-ray flux with hardness ratio (bottom panel in each sub-figure) for the four states identified in the 
$\gamma$-ray light curve. }
\label{flux_alp_hr}
\end{figure*}

\begin{figure}
\includegraphics[height=7.5cm,width=9cm]{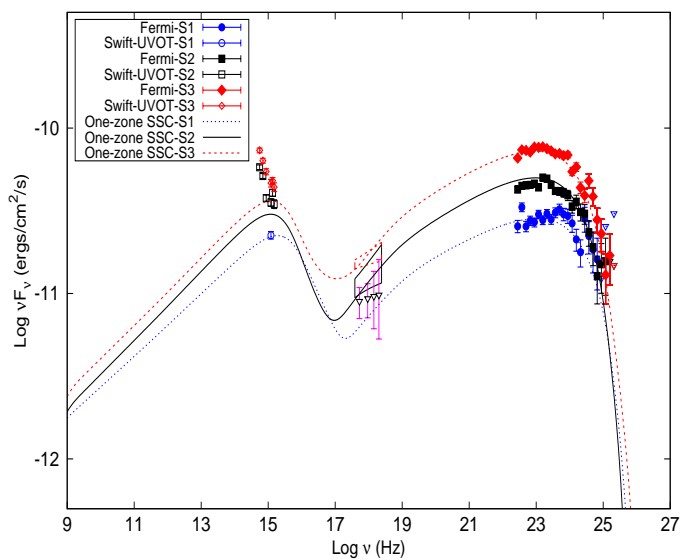}
\caption{SED of NGC 1275 during different activity states. 
The dotted, solid and dashed lines represents the total (Sync+SSC) SED model contribution from S1, S2, and S3 respectively. 
The bow-ties in X-ray band for S1, S2, and S3 are from {\sl Swift}-XRT. The downward triangles are for upper limit values.}
\label{sed_3states}
\end{figure}

\begin{figure*}
\includegraphics[height=6cm,width=8cm]{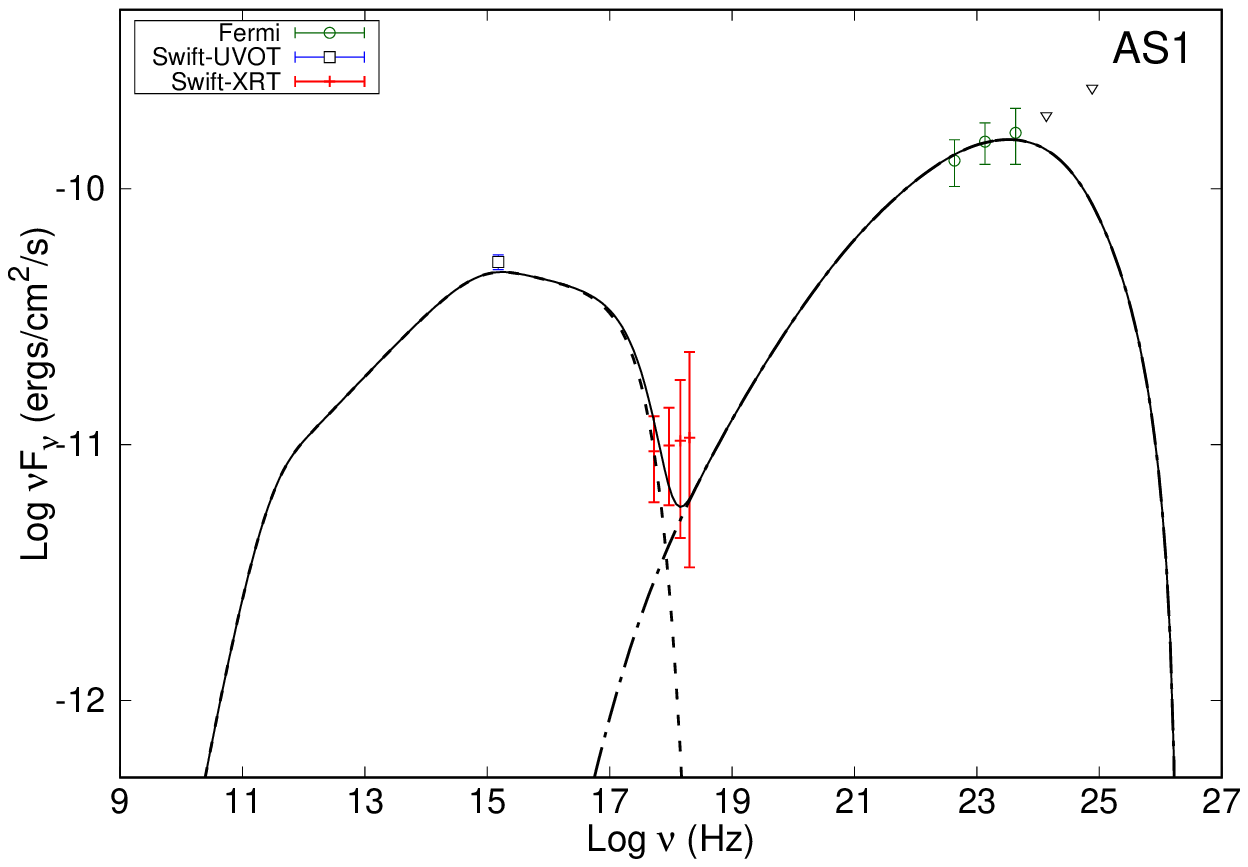}
\includegraphics[height=6cm,width=8cm]{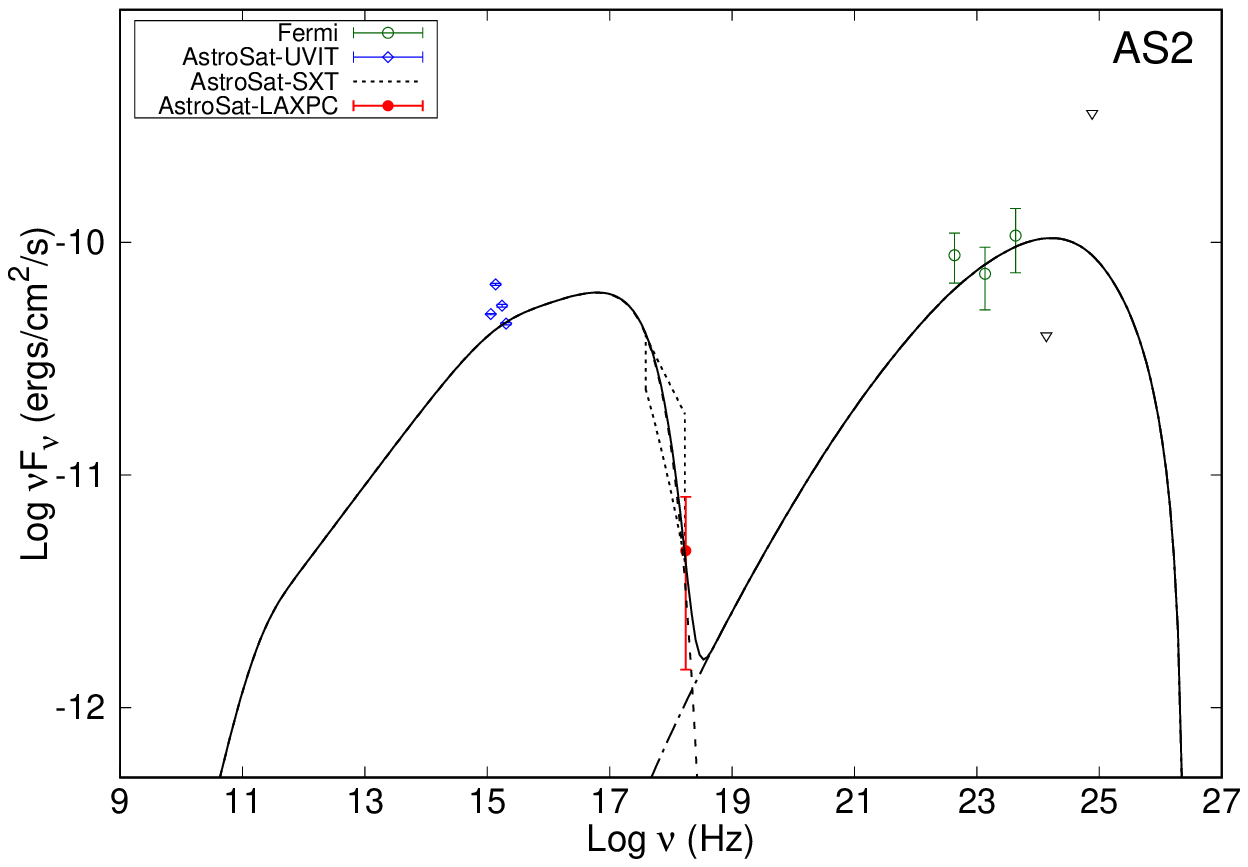} \\

\caption{SED of NGC 1275 during {\sl AstroSat} observing periods. 
The dashed line represents the contribution from synchrotron emission and 
dash-dotted line represents the contribution from SSC emission.
The solid line represents the total (Sync+SSC) SED model contribution. 
The downward triangles are for upper limit values.}
\label{sed_astrosat}
\end{figure*}

\section{Discussion}
\subsection{Temporal Behaviour}

Similar to the long-term variability as observed in blazars \citep[e.g.,][]{bhattacharya2013,Bhattacharya2017,Bhattacharya2021}, 
\citet{ngc1275_fermi} noticed that the $\gamma$-ray flux observed 
during the first few months of {\sl Fermi} observations was nearly a 
factor of ten higher than the EGRET flux upper limit suggesting 
a presence of flux variability 
of a much longer timescale in NGC 1275. 
\citet{dutson2014} reported the presence of a decade timescale variability in this source. 
Based on the first 
year of {\sl Fermi} observations of NGC 1275, \citet{kataoka2010} noticed a monthly 
timescale variability of the flux and the spectral indices. 
Short-term variability on hours and sub-week timescales was 
also noticed in this source \citep{Baghmanyan2017, ngc1275_fermi_2011}.
The hour-timescale variability reported by \citet{Baghmanyan2017} 
suggests an extremely compact emitting region.  
Short term ($\sim$ hours) variability 
in the optical band was also noticed in a few other $\gamma$-ray detected 
misaligned active galaxies \citep{Bhattacharya2019}.

\citet{Tanada2018} studied the $\gamma$-ray variability 
of NGC 1275 utilising $\sim 8$ years of {\sl Fermi} observations and
defined two epochs (epoch A and epoch B) 
in the light curve based on the fluctuations in the 
spectral indices. 
In the present study, we 
have defined $4$ activity states 
based on the increase in the baseline flux nature noticed in the 
monthly averaged $\gamma$-ray light curve of NGC 1275 as described in Section~\ref{temporal}. 
As represented in the top panels of Fig.~\ref{flux_alp_hr}, no significant 
correlation was noticed between the monthly averaged $\gamma$-ray 
flux and the spectral parameter $\alpha$ during S1, S2, and S3 states 
with Pearson correlation coefficient value $0.3$, $-0.2$, and $-0.3$ respectively. 
However, during S4 state, a hint of ``flatter when brighter'' 
feature was noticed with a correlation coefficient of $\sim -0.8$.

The hardness ratio of this source was also calculated in each monthly bin 
utilising the monthly averaged $\gamma$-ray fluxes 
in $100$ MeV - $1$ GeV and $1$ GeV - $100$ GeV energy bands, 
which is represented in bottom panels of Fig.~\ref{flux_alp_hr}. 
Similar to the findings of the flux-spectral parameter correlation study, no 
significant correlation was noticed 
between the flux and the hardness ratio values during S1, S2, and S3 
states with the Pearson correlation coefficient 
value $-0.3$, $0.1$, and $0.2$, respectively. 
However, a significant correlation was found during S4 with 
Pearson correlation coefficient $0.6$,  
supporting a `harder when brighter' scenario. 
\citet{Tanada2018} found a correlation between the $\gamma$-ray 
flux and the hardness ratio during one of the flares in epoch A of their study. 
However, in epoch B of their study, which overlaps with S2 and S3 states of this work, no 
significant correlation was noticed. 

A strong positive correlation between the optical and the $\gamma$-ray 
fluxes was noticed by \citet{Aleksic2014} between October 2009 to 
February 2011. \citet{Fukazawa2018} reported the first 
positive correlation between fluxes in the X-ray and the $\gamma$-ray energy bands 
during $2013-2015$. In this work, we noticed a correlated increase 
of $\gamma$-ray, X-ray and UV/optical fluxes  
averaged over S1, S2 and S3 states 
without any appreciable change in the spectral indices. 

\subsection{Modelling of Spectral Energy Distribution} 
Usually one zone synchrotron self Compton (SSC)
jet model has been used to 
explain the broadband 
SEDs of NGC 1275 \citep[e.g.,][]{ngc1275_fermi,
kataoka2010,Aleksic2014,Tanada2018}. 
However, \citet{tavecchio2014} considered a ``spine-layer'' scenario 
to explain the broadband SED of this source during the 
MAGIC campaign \citep{Aleksic2014}. 

From the broadband SED modelling of the $\gamma$-ray flaring and quiescent states of NGC 1275, 
\citet{Tanada2018} suggested that flux changes in epoch A were caused by the 
injection of the high-energy electrons in the jet, while a change of Doppler factor 
could explain the observed flux variations in epoch B. 
\citet{Aleksic2014} derived the parameters,  
which are in the typical range found for BL Lacs (except for bulk Lorentz factor),  
from the broadband SED modelling of 
simultaneous observations in two campaigns (October 2009-February 2010 
and August 2010-February 2011) of NGC 1275.  
They suggested that NGC 1275 could be a misaligned BL Lac with large jet inclination angle and small  
bulk Lorentz factor. Alternatively, it might be more aligned with smaller jet inclination 
angle and higher bulk Lorentz factor.

\begin{table*}
\centering
\caption{ Model parameters for the SED}
\label{table8}
\begin{tabular}{lccccccr} 
\hline
Parameter         &Symbol     &State 1    &State 2      &State 3    &{\sl AstroSat}-AS1   &{\sl AstroSat}-AS2 \\
\hline
Minimum electron Lorentz factor &$\gamma_{1}$ &  $15.0$ &  $19.0$      &   $20.0$     &  $450$       & $350$ \\
Maximum electron Lorentz factor &$\gamma_{2}$ &$2.6\times10^{5}$  &$2.0\times10^{5}$  &$4.0\times10^{5}$  &$5.7\times10^{5}$  &$7.1\times10^{5}$  \\
Break Lorentz factor &$\gamma_{b}$ &$6.0\times10^{4}$ &$5.0\times10^{4}$   &$5.0\times10^{4}$ &$3.0\times10^{4}$ &$4.0\times10^{4}$   \\
Normalisation of particle spectrum &$N_{0}$ & $3.25\times10^{44}$ &$7.93\times10^{44}$  &$9.31\times10^{44}$ &$4.90\times10^{45}$  &$1.81\times10^{45}$     \\
Particle spectral index (before break)  &$p_{1}$  & $2.6$   &   $2.6$     &  $2.6$     &  $2.5$       & $2.3$ \\
Particle spectral index (after break)  &$p_{2}$  &  $4.1$ &     $4.5$   &   $4.4$     &   $3.1$      & $2.8$ \\

\hline
\end{tabular}
\\

\end{table*}

In the present work, 
we have studied the average behaviour of the source at its various activity states.  
Broadband SEDs of the source were constructed and modelled  
during S1, S2 and S3 states. 
Due to the lack of adequate multi-band data, averaged broadband 
SED study could not be carried out during the S4 state. 
We also constructed and modelled the broadband SEDs during 
AS1 and AS2 states. 
During AS1, data 
from {\sl UVW2} filter of {\sl Swift}-UVOT, 
{\sl Swift}-XRT and {\sl Fermi}-LAT were used 
to model the broadband SED.   
During AS2, data from 
four filters of {\sl AstroSat}-UVIT, {\sl AstroSat}-SXT, 
{\sl AstroSat}-LAXPC and {\sl Fermi}-LAT were used 
to model the broadband SED.  
We considered a single zone leptonic model 
that includes synchrotron and SSC 
emission by non-thermal relativistic electrons in relativistically moving 
emission region (blob) in AGN jet, as described in \cite{bhattacharyya2018}. 
The energy distribution of relativistic jet electrons was considered 
as a broken power-law, which is given by
\begin{eqnarray}
	N(\gamma) &= & N_0 \Big(\frac{\gamma}{\gamma_b}\Big)^{-p_1} \;\;\;\; \mbox{ for  } \gamma_1 \le \gamma \le \gamma_b \nonumber \\
                  &=& N_0 \Big(\frac{\gamma}{\gamma_b}\Big)^{-p_2} \;\;\;\;  \mbox{ for  } \gamma_b \le \gamma \le \gamma_2 \nonumber
\end{eqnarray}

where, $\gamma_1$, $\gamma_2$, and $\gamma_b$ are minimum, 
maximum and break Lorentz factors whereas $p_1$ and $p_2$ are the particle spectral indices
before and after the break Lorentz factor.

For modelling of SEDs in all the activity states, the employed 
jet inclination angle ($\theta$) was $20^{\circ}$, consistent with previous studies on broadband SED of this source  
\citep[ e.g.,][]{ngc1275_fermi, Aleksic2014, Tanada2018, Fukazawa2018}.
However, this value of $\theta$ is much smaller than that inferred from radio observations 
($\theta = 30^{\circ}-55^{\circ}$; \citealt{Walker1994} and 
$\theta = 49^{\circ}-81^{\circ}$; \citealt{fujita2017}). 
Other derived physical parameters for all the activity states are the blob radius of $R = 0.14\times10^{18}$ cm, 
the magnetic field of $B=0.08$ G and the bulk Lorentz factor $\Gamma=3.0$.  
These values are in the range considered in previous studies 
\citep[e.g.,][]{ngc1275_fermi, Aleksic2014, Tanada2018, Fukazawa2018}.

While modelling the S1, S2 and S3 states, the maximum value of the particle Lorentz factor and the break Lorentz 
factors were grossly estimated from the observed photon spectrum so that the 
maximum observed photon frequency and the spectral turn over in the IC hump 
could be generated. The minimum Lorentz factor was adjusted so that the SSC 
process could explain the X-ray flux. Nevertheless, with the given 
data none of these parameters could be constrained well for S1, S2 and S3 states. 

While modelling of the SEDs during AS1 and AS2, the primary approach 
followed to estimate the spectral parameters remained the same. However, 
for both the states, the soft and hard X-ray spectra appeared in the 
falling edge of the synchrotron peak. Therefore, the highest frequency 
of the synchrotron peak and the approximate turnover frequency of the 
SSC peak were used to estimate the maximum and the break Lorentz factors 
of the particle spectrum. Nevertheless, they could not be constrained well. 
Since the spectra moved towards higher energies, the minimum Lorentz factor 
of the electron distributions was pushed to higher values as compared to the 
states S1 - S3. The values of the minimum Lorentz factor of electrons 
comparable to what was obtained for states S1 - S3 would result in more radiation 
power in the lower frequency region. Therefore, the orders of magnitude of the 
Lorentz factors that we obtained from the modelling of the SEDs are relatively 
consistent, even though they are not fully constrained. 

The fitted SEDs during S1, S2, and S3 
states are shown in Fig~\ref{sed_3states}. SEDs 
during {\sl AstroSat} observing periods are shown in Fig~\ref{sed_astrosat}. 
Due to large error in X-ray flux and photon index values during S1 and AS1 states, 
we used the derived flux in four bands $1.6-3.0$, $3.0-5.0$, $5.0-7.0$ and $7.0-10.0$ keV 
using the ``cflux'' routine in \texttt{XSPEC} to construct X-ray SED 
as shown in Fig~\ref{sed_3states} and Fig~\ref{sed_astrosat}.
The results of our SED fitting are  
given in Table~\ref{table8}.

Our findings from the SED modelling of S1, S2, S3, AS1 and AS2 states are given below.   
\begin{itemize}

\item The optical spectral indices during S2 and S3 are steeper 
than that of X-ray, and $\gamma$-rays. Also, the observed flux in 
the optical band is significantly higher than the predicted flux 
from the synchrotron emission of the jet electrons. The observed 
nature of optical flux and spectra suggests that the optical emission 
might have originated external to the jet, probably from the accretion 
		disk/BLR, host galaxy as well as from the cluster. 
		  
	\item The $\gamma$-ray emission was well 
		explained by the SSC emission. 
		The UV emission is marginally higher than that predicted by the synchrotron emission process.
\item During S1, S2, and S3 states, the flux in the X-ray band was explained by 
the SSC emission. However, during AS1 and AS2, X-ray emission was explained by 
the synchrotron process.

\item An increase in jet particle normalisation 
with no significant variation in other parameters was also noticed 
during S1, S2, and S3 states. While studying long-term X-ray behaviour 
of NGC 1275 during 2006-2015, \citet{Fukazawa2018} speculated that 
increase in the electron density could be one of the possible explanations of 
		the observed long-term X-ray flux increase. 

\item Unlike S1, S2, and S3 states where energy distribution of 
jet electrons was very close to a single power-law, 
a broken power energy distribution of jet electrons, with a much flatter 
spectrum, was noticed during both AS1 and AS2 observing periods. 
Also, a significant increase in the minimum and maximum electron Lorentz factors 
was noticed during {\sl AstroSat} observing periods.  
		The SEDs of the source during AS1 and AS2 show 
		significant changes as compared to the SEDs during S1, 
		S2 and S3 states. First, the overall luminosity of the
source increased. Second, the X-ray spectra, as obtained from XRT, SXT 
		and LAXPC, changed the slope during both AS1 and AS2 as 
		compared to the S1, S2 and S3 states. Finally the peak 
		of the SED in synchrotron as well as Compton hump shifted 
		to the higher energies in AS1 and AS2. Therefore the slope of the 
		X-ray spectra during {\sl AstroSat} observing periods indicated that it was part of the 
		synchrotron hump. To model such features in the SEDs of 
		AS1 and AS2 states, it was necessary to increase the values of
$\gamma_1$ and $\gamma_2$. The values of $\gamma_2$ were constrained 
		by both the X-ray and the $\gamma$-ray spectra.

\end{itemize}

\section{Conclusion} 

In this work, based on the variation in the $\gamma$-ray 
baseline flux, we identified four activity states of this source. 
We also observed the source twice with {\sl AstroSat} during its high 
$\gamma$-ray activity state. We found three 
distinct states (S1, S2, S3) with increase in the $\gamma$-ray 
baseline flux for $\sim 9$ years, followed by 
another state (S4) characterised by a large 
long term $\gamma$-ray flare. We determined the boundaries 
of these activity states by fitting linear functions to the 
cumulative flux distribution. A correlation study between 
the $\gamma$-ray flux and the spectral nature of the source was carried out. 
Also, broadband SEDs were constructed 
and modelled utilising observations from 
{\sl Fermi}-LAT, {\sl Swift} and {\sl AstroSat}. 
We conclude: 
 
\begin{itemize}
\item An increase of the $\gamma$-ray baseline flux with no appreciable 
change in averaged spectral properties was noticed 
during S1 to S3 state. Similar behaviour was also 
noticed in the UV and X-ray bands. 
An increase in the jet particle normalisation could explain this observed feature. 

\item No significant correlation was noticed 
between the $\gamma$-ray flux and the spectral parameter/hardness ratio during S1, S2, and S3 states. 
However, a hint of correlation was noticed during the S4 state. 
 
\item Based on the first two years of {\sl Fermi} observations \citet{ngc1275_fermi_2011} reported 
that ``NGC 1275 appeared to migrate from the FR I radio galaxy 
to the BL Lac object  region'' during large $\gamma$-ray flare. 
While explaining the steeper X-ray  
spectrum during 2010 flare of this source, \citet{Fukazawa2018} proposed 
that the X-ray emission might have synchrotron origin unlike SSC in normal state. 
In this work, we found that during S1, S2, and S3 states, which represent 
the long term averaged behaviour of this source, X-ray emission was 
well explained by the SSC process. However, during AS1 and AS2 observing periods,  
		there is an evidence of an increase in the synchrotron peak frequency 
		and the X-ray emission is explained by the synchrotron emission of jet electrons.
\end{itemize}
Long term study of NGC 1275 in this work provides a 
better understanding of {the underlying emission mechanism during various 
activity states.

\section*{Acknowledgements}
We thank the anonymous referee for his/her constructive comments
that helped us to improve the manuscript considerably. 
The author(s) thank Ranjeev Misra for discussions regarding {\sl AstroSat}-LAXPC and overall X-ray analysis, 
and Gulab Dewangan for discussion regarding {\sl AstroSat}-SXT analysis. 
The author(s) thank Jayashree Roy, Bitan Ghosal, Anil Tolamatti, and Ashish Devaraj (UVIT-POC) for their 
useful discussions. 
The author(s) acknowledge the financial support of 
Indian Space Research Organisation (ISRO) under
{\sl AstroSat} archival Data utilization program.
This publication uses the data 
from the {\sl AstroSat} mission of the ISRO, 
archived at the ISSDC. 
This work has been performed utilising the calibration data-bases and auxiliary analysis 
tools developed, maintained and distributed by {\sl AstroSat}-SXT team with members from various 
institutions in India and abroad. 
This work has made use of public
{\sl Fermi}-LAT data obtained from the {\it Fermi} Science Support Center
(FSSC), provided by NASA Goddard Space Flight Center. 
This research has made use of the NASA/IPAC Extragalactic Database (NED), 
which is operated by the Jet Propulsion Laboratory, California Institute of Technology, 
under contract with the National Aeronautics and Space Administration. 
This research has made use of data and/or software provided by the High 
Energy Astrophysics Science Archive Research Center (HEASARC),
which is a service of the Astrophysics Science Division at NASA/GSFC and the High 
Energy Astrophysics Division of the Smithsonian Astrophysical Observatory. 
This research has made use of the XRT Data Analysis Software (\textsc{xrtdas}) 
developed under the responsibility of the ASI Science Data Center (ASDC), Italy.
Manipal Centre for Natural Sciences, Centre of Excellence, Manipal Academy
of Higher Education (MAHE) is acknowledged for facilities and support.

\section*{Data availability} 
This work has made use of public
{\sl Fermi}-LAT data available at 
\url{https://fermi.gsfc.nasa.gov/cgi-bin/ssc/LAT/LATDataQuery.cgi}. 
This research has made use of data and software provided by the High 
Energy Astrophysics Science Archive Research Center (HEASARC) 
available at \url{https://heasarc.gsfc.nasa.gov/docs/software/lheasoft/}, 
{\sl Swift} data available at \url{https://heasarc.gsfc.nasa.gov/cgi-bin/W3Browse/w3browse.pl} 
and the NASA/IPAC Extragalactic Database (NED)}. 
This publication has also made use of the data   
from the {\sl AstroSat} mission of the ISRO, archived at the ISSDC 
(\url{https://astrobrowse.issdc.gov.in/astro\_archive/archive/Home.jsp}).
{\sl AstroSat} data will be shared on request to the corresponding author with the 
permission of ISRO.


\def\apj{ApJ}%
\def\mnras{MNRAS}%
\def\aap{A\&A}%
\def\apjl{ApJ}
\def\aj{AJ}
\def\physrep{PhR}
\def\apjs{ApJS}
\def\pasa{PASA}
\def\pasj{PASJ}
\def\nat{Natur}
\def\apss{Ap\&SS}
\def\araa{ARA\&A}
\def\aaps{A\&AS}
\def\ssr{Space Sci. Rev.}
\def\pasp{PASP}
\def\na{New Astron.}
\def \procspiekp{Space Telescopes and Instrumentation 2014: Ultraviolet to Gamma Ray}
\def \procspiekumar{Space Telescopes and Instrumentation 2012: Ultraviolet to Gamma Ray}
\def \procspieyadav{Space Telescopes and Instrumentation 2016: Ultraviolet to Gamma Ray}

\bibliographystyle{mnras}
\bibliography{ref}





\bsp	
\label{lastpage}
\end{document}